\documentclass[useAMS,usenatbib]{mn2e}
\usepackage{color}
\usepackage{array} 
\usepackage{amsmath}
\usepackage{graphicx}
\usepackage{epstopdf}
\usepackage[labelfont=bf,labelsep=period,justification=raggedright,singlelinecheck=false,font=small]{caption}
\usepackage{wrapfig}
\usepackage{float}
\usepackage{amssymb}
\usepackage[normalem]{ulem}
\usepackage{hyperref}
\usepackage{undertilde}
\usepackage[T1]{fontenc}
\usepackage[utf8]{inputenc}
\def\apj{ApJ}
\def\aap{A\& A}
\def\apjl{ApJL}
\def\apjs{ApJS}
\def\mnras{MNRAS}

\def\aj{AJ}

\def\nat{Nature}

\def\rmxaa{RMxAA}

\newcommand\altaffilmark[1]{$^{#1}$}
\newcommand\altaffiltext[1]{$^{#1}$}

\newcommand{\msun}{{\rm M}_{\odot}}

\restylefloat{table}

\begin{document}

\title[UDGs in FIRE]{The origin of ultra diffuse galaxies: stellar feedback and quenching}

\author[T. K. Chan et al.]
  {T. K. ~Chan\altaffilmark{1} \thanks{Email: (TKC) tkc004@physics.ucsd.edu},
  D. ~Kere\v{s}\altaffilmark{1} \thanks{Email: (DK) dkeres@physics.ucsd.edu},
  A. Wetzel\altaffilmark{2,3,4}\thanks{Caltech-Carnegie Fellow},
  P. F.~Hopkins\altaffilmark{2},
  C.-A. Faucher-Gigu{\`e}re\altaffilmark{5}, \newauthor
  K. El-Badry\altaffilmark{6},
  S. Garrison-Kimmel\altaffilmark{2},
  M. Boylan-Kolchin\altaffilmark{7}\\
  \altaffiltext{1}{Department of Physics, Center for Astrophysics and Space Sciences,University of California at San Diego,}\\ {9500 Gilman Drive, La Jolla, CA 92093, USA}\\
  \altaffiltext{2}{TAPIR, Mailcode 350-17, California Institute of Technology, Pasadena, CA 91125, USA} \\
\altaffiltext{3}{The Observatories of the Carnegie Institution for Science, Pasadena, CA 91101, USA} \\
\altaffiltext{4}{Department of Physics, University of California, Davis, CA 95616, USA} \\
\altaffiltext{5}{Department of Physics and Astronomy and CIERA, Northwestern University, 2145 Sheridan Road, Evanston, IL 60208, USA} \\ 
\altaffiltext{6}{Department of Astronomy and Theoretical Astrophysics Center, University of California Berkeley, Berkeley, CA 94720, USA}\\
\altaffiltext{7}{Department of Astronomy, The University of Texas at Austin, 2515 Speedway, Stop C1400, Austin, TX 78712-1205, USA}}

\maketitle

\begin{abstract}
We test if the cosmological zoom-in simulations of isolated galaxies from the FIRE project 
reproduce the properties of ultra diffuse galaxies (UDGs). We show that outflows that dynamically heat galactic stars, together with a passively aging stellar population after imposed quenching, naturally reproduce the observed population of red UDGs, without the need for high spin halos, or dynamical influence from their host cluster. We reproduce the range of surface brightness, radius and absolute magnitude of the observed red UDGs by quenching simulated galaxies at a range of different times.  They represent a mostly uniform population of dark matter-dominated dwarf galaxies with  $M_*\sim 10^{8}\,\msun$, low metallicity and a broad range of ages; the more massive the UDGs, the older they are. The most massive red UDG in our sample ($M_*\sim 3\times 10^8 \msun$) requires quenching at $z\sim 3$ when its halo reached $M_{\rm h}\sim 10^{11}\,\msun$.
Our simulated UDGs form with normal stellar-to-halo ratios and match the central enclosed masses and the velocity dispersions of the observed UDGs. 
Enclosed masses remain largely fixed across a broad range of quenching times because the central regions of their dark matter halos complete their growth early. 
If our simulated dwarfs are not quenched, they evolve into bluer low-surface brightness galaxies with M/L similar to observed field dwarfs. While our simulation sample covers a limited range of formation histories and halo masses,
we predict that UDG is a common, and perhaps even dominant, galaxy type around $M_{*}\sim 10^{8}\,\msun$, both in the field and in clusters. 
\end{abstract}


\label{firstpage}

\section{Introduction}
 Low surface brightness galaxies (LSBs) with large effective radii were detected and studied by many authors over the past several decades  \citep{Impe88,Both91,Dalc97,Cald06,McCo08}. \cite{vanD15} sparked a recent interest in LSBs by finding many LSBs in the Coma cluster, named ultra diffuse galaxies (UDGs), with absolute magnitudes comparable to those of dwarf galaxies ($M_g\sim-14$), but with effective radii as large as the Milky Way (MW) ($\sim 4\,{\rm kpc}$) and surface brightnesses of $\sim 25 \,{\rm mag/arcsec^2}$. They appear spheroidal and red, indicating old stellar populations. 
 
 Since then, a large number of UDGs have been discovered in the Coma cluster \citep{Koda15}, the Virgo cluster \citep{Cald06,Miho15}, the Fornax cluster \citep{Muno15}, clusters with $z\sim 0.044-0.063$ \citep{vande16}, the Abell 2744 cluster \citep{Jans17}, the Abell S1063 clusters \citep{Lee17}, the Pisces--Perseus Supercluster \citep{Mart16}, the M77 group \citep{Truj17}, the elliptical galaxy NGC 5485 \citep{Merr16}, three nearby isolated groups \citep{Roma17b} and the Hickson Compact Group 95 \citep{Shi17}.

Because of their large effective radii and low inferred stellar masses, \cite{vanD15} proposed that UDGs are ``failed'' ${\rm L}\star$ galaxies initially forming in relatively massive halos that were quenched at $z\sim 2$. This hypothesis is supported by the stellar velocity dispersion and the number of globular clusters (GCs) of a massive UDG, Dragonfly 44, in the Coma cluster for which \citet{vanD16,vanD17} inferred a total halo mass $\sim 10^{12}\msun$. 

However, from recent observations of the GC systems of other UDGs,  \cite{Beas16a, Beas16b, Peng16} argued that the UDGs are ``failed'' dwarf galaxies.  By measuring the velocity dispersion of the GC system in a UDG (VCC 1287 in Virgo Cluster), \cite{Beas16a} inferred a dynamical mass of $4.5 \times 10^{9} \msun$ within 8.1 kpc. By comparing its dynamical mass with numerical simulations, they estimated its halo mass $M_{200} = (8\pm 4)\times 10^{10}\msun$, comparable to a dwarf galaxy halo.  By measuring the number of GCs in Coma UDGs and assuming the correlation between GC number and halo mass, \cite{Amor16b} found most of the Coma UDGs reside in dwarf halos. Similar conclusions were reached with measurements of GC specific frequencies of UDGs \citep{Beas16a, Beas16b, Peng16}. Furthermore, \cite{Roma17a} revealed that the spatial distribution of UDGs in a galaxy cluster resembles the distribution of dwarf galaxies rather than ${\rm L}\star$ galaxies. Based on these measurements, \cite{Beas16b} argued that UDGs are quenched galaxies that inhabit Large Magellanic Cloud (LMC)-sized halos and quench their star formation at $z\sim 3$. In this scenario, cluster UDGs have to be quenched for more than 10 Gyr. \cite{Sifo17} used weak gravitational lensing to show that the average virial mass of 784 UDGs in 18 clusters is $m_{200}\leq 10^{11.80}\msun$, consistent with dwarf halo masses but leaving a possibility of the most massive UDGs to be hosted in MW-mass halos. 

\cite{Yozi15} similarly argued that UDGs have dwarf progenitors, but they quenched at much later times. They simulated interactions between a cluster and an infalling diffuse dwarf galaxy at $z\sim 2$, and showed that the harsh cluster environment can rapidly halt any ongoing star formation. Their initial conditions assumed the infalling dwarf was hosted in a high spin halo, allowing the galaxy to be much more diffuse than normal galaxies even before interacting with the host cluster.  Following this line of thought, \cite{Amor16} proposed that UDGs are the high spin tail of the dwarf galaxy population, so they are diffuse even without interacting with the cluster. They predicted there should also be a field population of UDGs but with possibly different morphologies and colors. \cite{Rong17} supported this hypothesis by finding that UDGs in their cosmological simulations reside in high spin halos. 

Upon finding UDGs with bluer color far from clusters, \cite{Roma17b} and \cite{Truj17} suggested that red UDGs in clusters might be initially low surface brightness diffuse dwarf galaxies born in the field that are later processed in groups and ultimately accreted into galaxy clusters.

In this paper, we use cosmological zoom-in simulations from the FIRE simulation to study the effects of stellar feedback and quenching on the progenitors of UDGs. Stellar feedback is known to shape dark matter (DM) profiles, creating large cores in the DM distribution of dwarf galaxy halos  \citep{Nava96,ElZa01,Gned04,Read05,Gove10,Pena12,Gove12,Pont12,Macc12b,Teys13,DiCi14,Pont14,Chan15,Toll16}. Feedback can also drive significant radial migrations of stars via two processes: (1) inflowing/outgoing gas clouds can form stars that inherit the velocities of the gas clouds and continue migrating within their first 100Myr; (2) Feedback-driven gas outflows modify central gravitational potential and transfer energy to stars non adiabatically (in the same manner as in DM core creation; see \citealt{ElBa16}). Through these processes, feedback expands galaxies into diffuse spheroids, producing large effective radii \citep{ElBa16} and large axis ratios \citep{Whee17} simultaneously. Effects of stellar feedback on both DM and stellar distributions peak at $M_* \sim 10^8\msun$, which is also a typical mass of the observed UDGs.

Using cosmological simulations of isolated galaxies with stellar feedback, \citet{DiCi17} recently also showed that feedback can produce extended stellar profiles similar to observed UDGs. Our study differs both in the stellar feedback model and in the inclusion of the effect of quenching, which has significant effects on the formation of red UDGs (see \S~\ref{implication} for a comparison of their findings with our work).

In \S~\ref{method} we describe the simulation methodology, the suite of simulations used in this paper and the method for mock observations with GALFIT. In \S~\ref{results}, we show how radius, surface brightness and other properties of simulated galaxies change with quenching time. In \S~\ref{discussion} we discuss the structural properties of our dwarfs, the connections to the formation scenarios discussed in the literature as well as the implications for the properties of field dwarf galaxies. Finally we present our conclusions in \S~\ref{conclusions}.

\section{Method}
\label{method}
\subsection{Simulation code and setup}
\label{sec:simulations}

\begin{table*}
\centering
    \begin{tabular}{llllllll}
    \hline\hline
    Name & $M^0_{\rm h}$&$M^0_{*}$ &$R_{\rm vir}$& $m_{\rm b}$ & $\epsilon_{\rm b}$ & $m_{\rm dm}$ & $\epsilon_{\rm dm}$                     \\
         & [$\msun$] & [$\msun$]&  [kpc]&[$\msun$] & [pc] & [$\msun$] &  [pc]  \\
    \hline
    	\hline
    {\bf m10z} & 3.5e10            &5.3e7    & 85                                &2.1e3  &0.43 & 1.0e4 & 43 \\
    {\bf m11a} &4.1e10&1.2e8& 90&2.1e3  &0.43  & 1.0e4 & 43\\
    {\bf m11b}&4.4e10&1.1e8& 92&2.1e3  &0.43  & 1.0e4 & 43\\
    {\bf m11q}  & 1.2e11                   & 1.0e9      &1.4e2                    & 7.1e3                                         & 0.42                                            & 3.5e4                                             & 14                                                     \\
    {\bf m11c} & 1.5e11            &2.0e9      & 1.4e2                               &1.7e4  &0.86 & 8.3e4                                       & 86   \\
    {\bf m11f} & 4.9e11            &2.7e10      & 2.1e2                               &1.7e4  &0.86 & 8.3e4                                       & 86   \\

    \hline  
    \end{tabular}

 	\caption{Simulation details. $M^0_{\rm h}$ and $M^0_{*}$ are the halo mass and stellar mass (within 0.2$R_{\rm vir}$) of the largest halo in the zoom-in region at $z=0$; $R_{\rm vir}$ is the virial radius; $m_{\rm b}$ is the mass of a gas particle in the simulation; $m_{\rm dm}$ is the mass of a DM particle in the simulation. $\epsilon_b$ is the minimum gravitational softening of a gas particle; $\epsilon_{\rm dm}$ is the Plummer equivalent gravitational softening of a DM particle.  All simulations are a part of the FIRE-2 simulation suite \citep{FIRE2}. The initial conditions for {\bf m11q} are from \citet{Kimm14}, while {\bf m11z} and {\bf m11c} are from \citet{Chan15}. {\bf m11a}, {\bf m11b} and {\bf m11f} are newly targeted halos in the mass range relevant for UDGs. Note that \citet{FIRE2} presented higher resolution runs of {\bf m10z} and {\bf m11c}, which we discuss in Appendix \ref{relimit}.} 
\label{SIC}
\end{table*}

Our simulations utilize the GIZMO\footnote{http://www.tapir.caltech.edu/$\sim$phopkins/Site/GIZMO} code \citep{Hopk15} in the mesh-free Lagrangian finite mass (MFM) mode for hydrodynamics. GIZMO uses an updated version of the PM+Tree algorithm from Gadget-3 \citep{Spri05} to calculate gravity and adopts fully conservative adaptive gravitational softening for gas \citep{Pric07}.  We employ the zoom-in technique to reach high resolutions in a cosmological environment and evolve simulations from $z=99$ to $z=0$.

Gas cooling is calculated with the tabulated cooling rates from CLOUDY \citep{Ferl13} for $T=10-10^{10}$ K, including atomic, metal-line, and molecular cooling. We apply the redshift-dependent ultraviolet background model from \cite{Fauc09} that ionizes and heats gas in an optically thin approximation and use an approximate prescription to account for self-shielding of dense gas. 

Star formation and stellar feedback are implemented using the FIRE-2 algorithm \citep{FIRE2}, which is an updated version of the FIRE feedback scheme from \citet{FIRE}.  Briefly, stars form in self-gravitating molecular gas at densities  $n_{\rm H} \geq 1000\, {\rm cm^{-3}}$, with 100\% instantaneous efficiency per free fall time.
Stellar feedback physics implemented includes stellar winds, radiation pressure from young stars, Type II and Type Ia supernovae, photoelectric heating, and photoheating from ionizing radiation. We calculate the energy, momentum, mass and metal return according to the STARBURST99 stellar population synthesis model \citep{Leit99} and Kroupa IMF \citep{Krou02}. Full details of the implementation of gas and gravitational physics are provided in \citet{FIRE2}.

All simulations analyzed in this work are a part of the FIRE-2 simulation suite of the FIRE project\footnote{http://fire.northwestern.edu}.  
Most are based on the initial conditions previously explored with FIRE-1 models in \citet{FIRE} and \citet{Chan15}\footnote{The FIRE-1 runs corresponding to {\bf m10z}, {\bf m11q} and {\bf m11c} were named as {\bf m10h573}, {\bf m11} and {\bf m11h383} respectively.}\footnote{We note that our values of $m_{\rm b}$ and $m_{\rm dm}$ for the runs included in \citet{Chan15} differ from values in Table 1 in that paper owing to their omission of  the factor of $h=0.7$. This omission did not affect any of the results quoted in that paper.}, while several additional galaxies were specifically simulated to explore the relevant mass scale of UDGs. Our sample includes all FIRE-2 galaxies with $z=0$ stellar mass $5\times 10^7 - 2\times 10^9 \msun$ and one additional higher mass galaxy, {\bf m11f}, selected to explore the UDG formation with quenching at high redshift. All satisfy the absolute magnitude range of the observed UDGs in at least one simulation snapshot in the redshift interval $z=0-3$ when post-processed with our mock-observations. Here we focus on properties of stellar population in our simulations; their gas properties were explored in \citet{ElBa17b}. Parameters of the simulated galaxies are listed in Table \ref{SIC}.

The galaxies we examine are isolated field dwarfs with ${M}_{\rm h}\sim 10^{10-11}\msun$ at $z=0$, where the effects of stellar feedback on the underlying density distributions are {\it large} (see e.g. \citealp{Chan15}, \citealp{ElBa16}, \citealp{Toll16}; but also \citealp{Onor15} and \citealp{Fitt17} for the effect in lower mass halos). Our simulations were run in a `standard' flat $\Lambda$CDM cosmology with the following cosmological parameters:  $\Omega_0\approx 0.27$, $\Lambda\approx0.73$, $\Omega_b\approx 0.045$ and  $h\approx0.7$.

We note that our simulated halos have a normal distribution of spin parameters. We measured the spin parameters of our DM halos (in default runs with full hydrodynamics and feedback) and found that at $z=0$ all except one are within 1$\sigma$ of the measured spin parameter distribution from \cite{maccio08} with values $\lambda \sim 0.02-0.035$. The exception is {\bf m11b} whose spin parameter is about 2$\sigma$ above the mean with spin parameter $\lambda=0.077$.

\subsection{Simulation analysis and mock observations}
\label{mock}

\begin{figure*}
\includegraphics[scale=0.4]{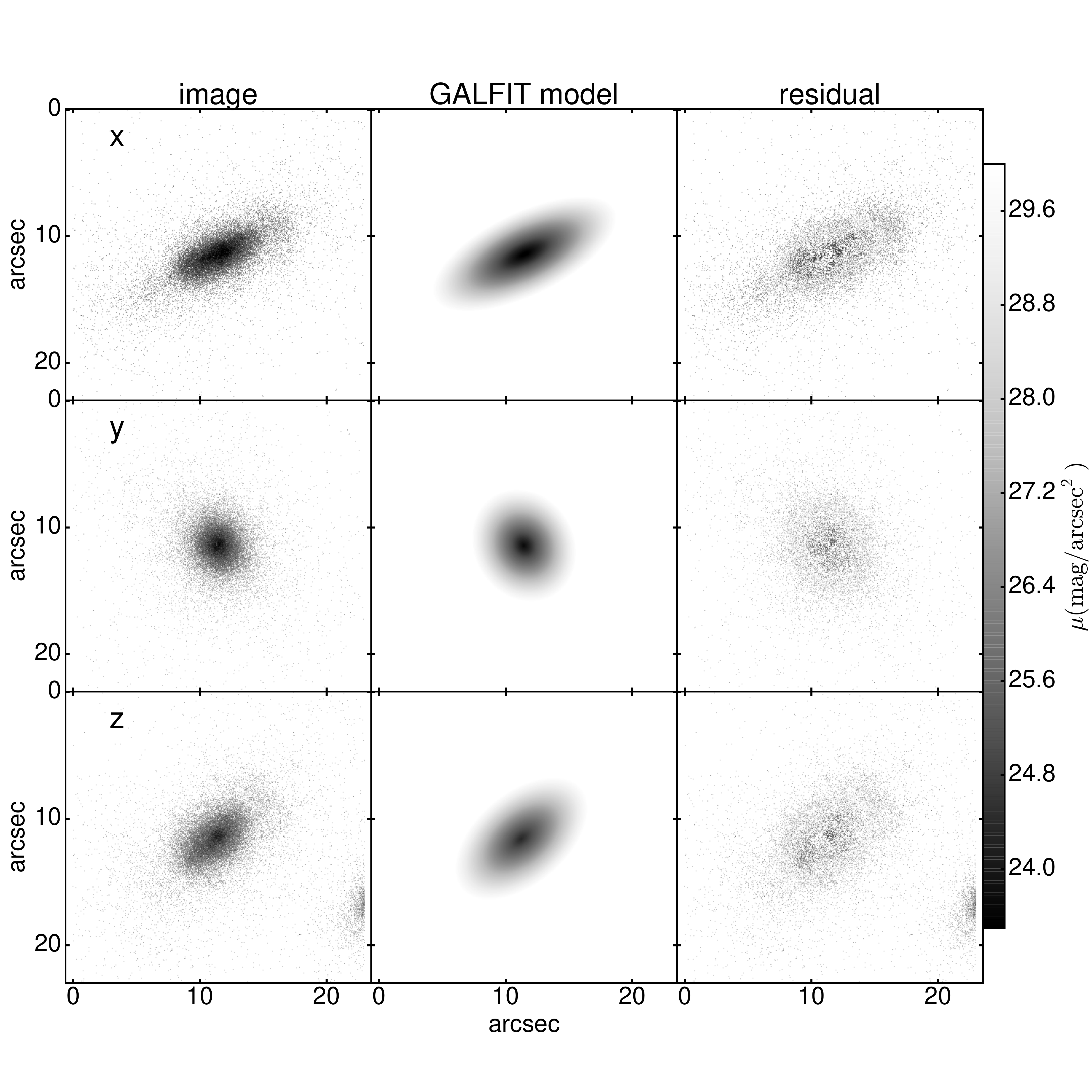}
\caption{The mock {\it g}-band image of {\bf m11b} ($ M_*\sim 10^8\msun$) quenched at a cosmic time $t_{\rm q}=6$ Gyr and passively aged stellar population from $t_{\rm q}$ to $z=0$, the best-fitting GALFIT model and the residuals from the fit. Each panel spans  16 kpc $\times$ 16 kpc ($23''\times 23''$ if we place the galaxy at the distance of the Coma cluster) and is shaded according to surface brightness. The top, middle and bottom panels show images viewed along x, y and z direction, respectively.}
\label{galfitsamplel}
\end{figure*}

All of our isolated dwarf galaxies remain gas-rich and star-forming until $z = 0$. This is in contrast to UDGs observed in galaxy clusters, many of which are quenched, likely because of the environmental effect of the clusters.  To mimic the quenching of star formation that likely occurs during the infall into the cluster environment, we artificially stop each galaxy's star formation at the assumed quenching time $t = t_{\rm q}$ (measured as cosmic time starting from the Big-Bang) and passively evolve its stellar populations to $z=0$. The minimum $t_{\rm q}$ we consider is 2 Gyr (i.e. $z\sim 3.5$) since all of the simulated galaxies in this work have sizes smaller than $\sim$1 kpc at earlier times and would therefore not satisfy our observationally motivated UDG selection (see \S~\ref{results}).

We assume that stellar morphology has not changed since $t_{\rm q}$. Even if quenching processes remove galactic gas (e.g. via ram pressure), this is a reasonable assumption as long as the galaxy is DM-dominated and stays far from the cluster center where cluster tidal interactions are important. In Appendix \ref{agasre} we study dynamical effects of gas removal and show that they tend to slightly decrease surface brightness but do not affect any of our conclusions. We do not attempt to account for other possible cluster interactions (e.g. tidal disruptions, galaxy harassment, etc.), which would require a full scale galaxy cluster simulation that is beyond the scope of our paper. In other words, our UDG candidates are simulated dwarf galaxies at $t_{\rm q}$, but with their stellar populations artificially aged to $z=0$.

In order to compare our simulated galaxies with observations, we produce mock images at $z=0$ for $t_{\rm q} \sim 2-14$ Gyr ($z_{\rm q} \sim 3.4-0$) with passively evolved stellar populations and perform mock observations to estimate their {\it g}-band surface brightnesses $\mu(g)$, effective radii $r_{\rm eff}$, and g-i colors. We follow the steps in \cite{vanD15} closely for a more direct comparison.  The galaxy images are initially centered on the halos of their main progenitors identified with the Amiga Halo Finder (AHF) \citep{Knol09}, which uses an adaptive mesh refinement hierarchy \citep{Kneb01} to define centers.  Halo centers may not coincide with galaxy centers during ongoing mergers or instabilities, however, so we relocated centers with a $\chi^2$ minimization on galaxy images with GALFIT \citep{Peng02}. To calculate enclosed masses (e.g. in Figures \ref{Menc} and \ref{Mencudg}), we applied the two-step procedure described in \cite{Chan15} to center on the stellar distribution of the galaxy.  We use AHF and virial overdensities from \cite{Brya98} to calculate virial mass $M_{\rm h}$, virial radius $R_{\rm vir}$ and $M_{*}$, the total stellar mass enclosed within $0.2R_{\rm vir}$. 

\cite{vanD15} inferred axis ratios and effective radii from combined g+i band images, and surface brightnesses from {\it g}-band images. To follow their procedure, we generated a table of SDSS (Sloan Digital Sky Survey) g and i band luminosities for stellar populations of different ages and metallicities\footnote{Although we do not use the Canada-France-Hawaii Telescope (CFHT) MegaCam filters, which the observations in \cite{vanD15} are performed with, the difference between the SDSS and CFHT filters are negligible and would not affect our main results.} with the Flexible Stellar Population Synthesis model (FSPS) \citep{Conr09}, assuming the latest Padova stellar evolution model \citep{Mari07,Mari08} and the Kroupa initial mass function \citep{Krou02}. The luminosity of each stellar particle is interpolated from the table according to their stellar ages, masses and metallicities. Then, for each simulation output we project the g + i band luminosities over a 40 kpc $\times$ 40 kpc (60 kpc $\times$ 60 kpc for {\bf m11f}) region onto $1000^2$ uniform mesh, and generate mock galaxy images. We explore the effects of lowering the image resolution (down to 100x larger pixels, i.e. 100x100 pixels per image) in Appendix \ref{appendgalfit}  and found that this does not significantly affect our results\footnote{This is because GALFIT attempts to fit for average surface brightness within each elliptical ring.}.

We also generate images with {\it g}-band luminosities to estimate {\it g}-band surface brightness. We do not account for any dust attenuation because we assume all gas is removed immediately after the infall (we briefly discuss the dust attenuation effects at $z=0$ in \S~\ref{implication}). The left panel of Figure \ref{galfitsamplel} shows the processed g+i band image of {\bf m11b} at $z=0$ with $t_{\rm q}=6$ Gyr (i.e. passively evolved from $z=1$), viewed along three perpendicular directions.

To estimate structural parameters from the mock images, we fit them with the Sersic profiles using GALFIT \citep{Peng02,Peng10}, similar to the techniques used in other UDG observations \citep[e.g.][]{Koda15,Miho15}. We allow $n_s$ to vary in our fits to increase convergence \citep{Koda15,Miho15}. Our galaxies have $n_s=0.8\pm0.4$, close to the $n_s=1$ profiles, used in \citet{vanD15}. We have compared central surface brightness of our galaxies obtained with the $n_s=1$ fits to those with variable $n_s$ and found only minor differences. Our fits do not account for sky noise: we have tested adding sky noise to our images in Appendix \ref{appendix:background} and again found very little difference in the inferred properties of our galaxies (see Figure \ref{musky}). The middle panels of Figure \ref{galfitsamplel} show the GALFIT models and the right panels show the residuals. 

\begin{figure*}
\includegraphics[scale=0.4]{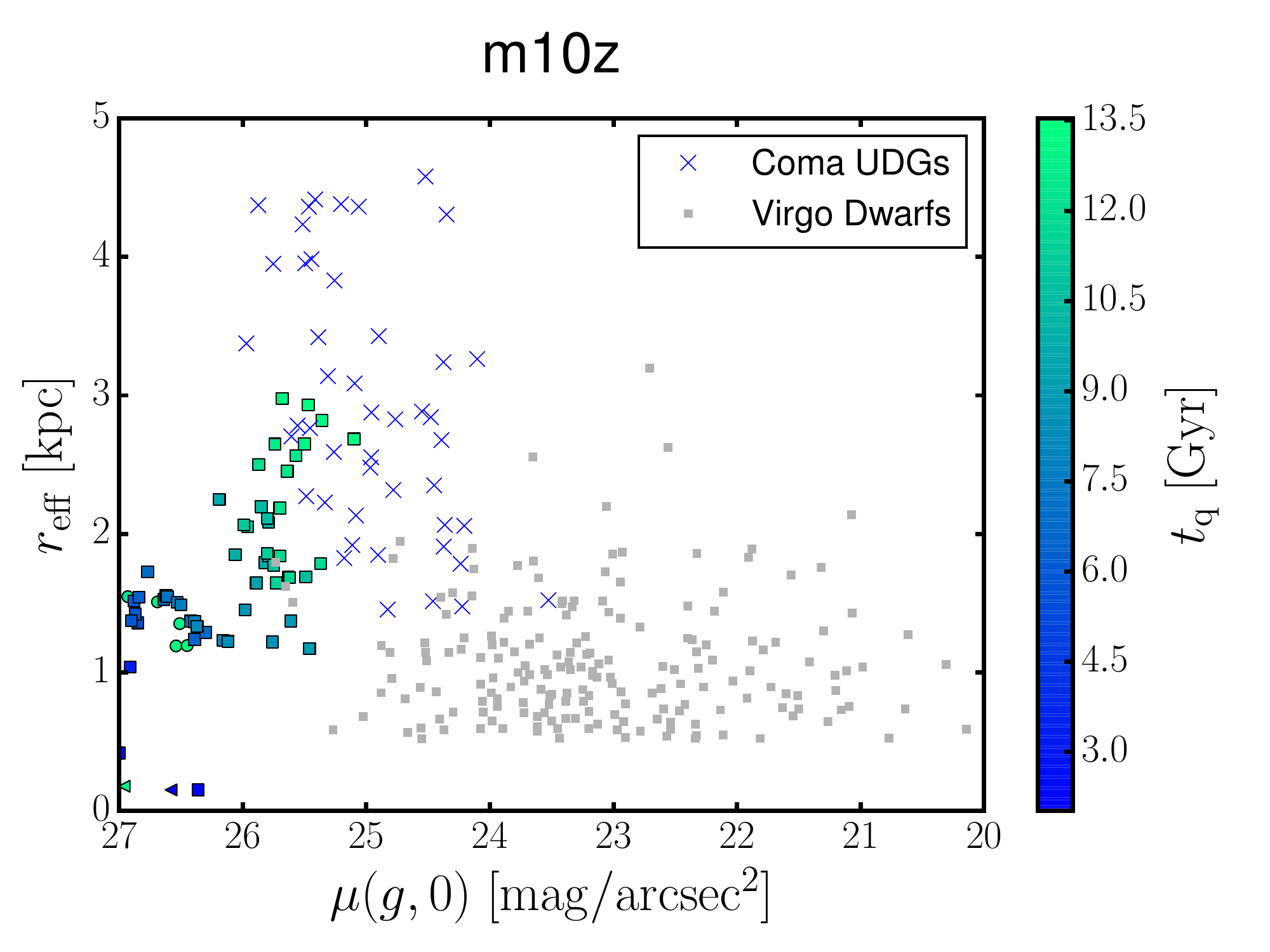}
\includegraphics[scale=0.4]{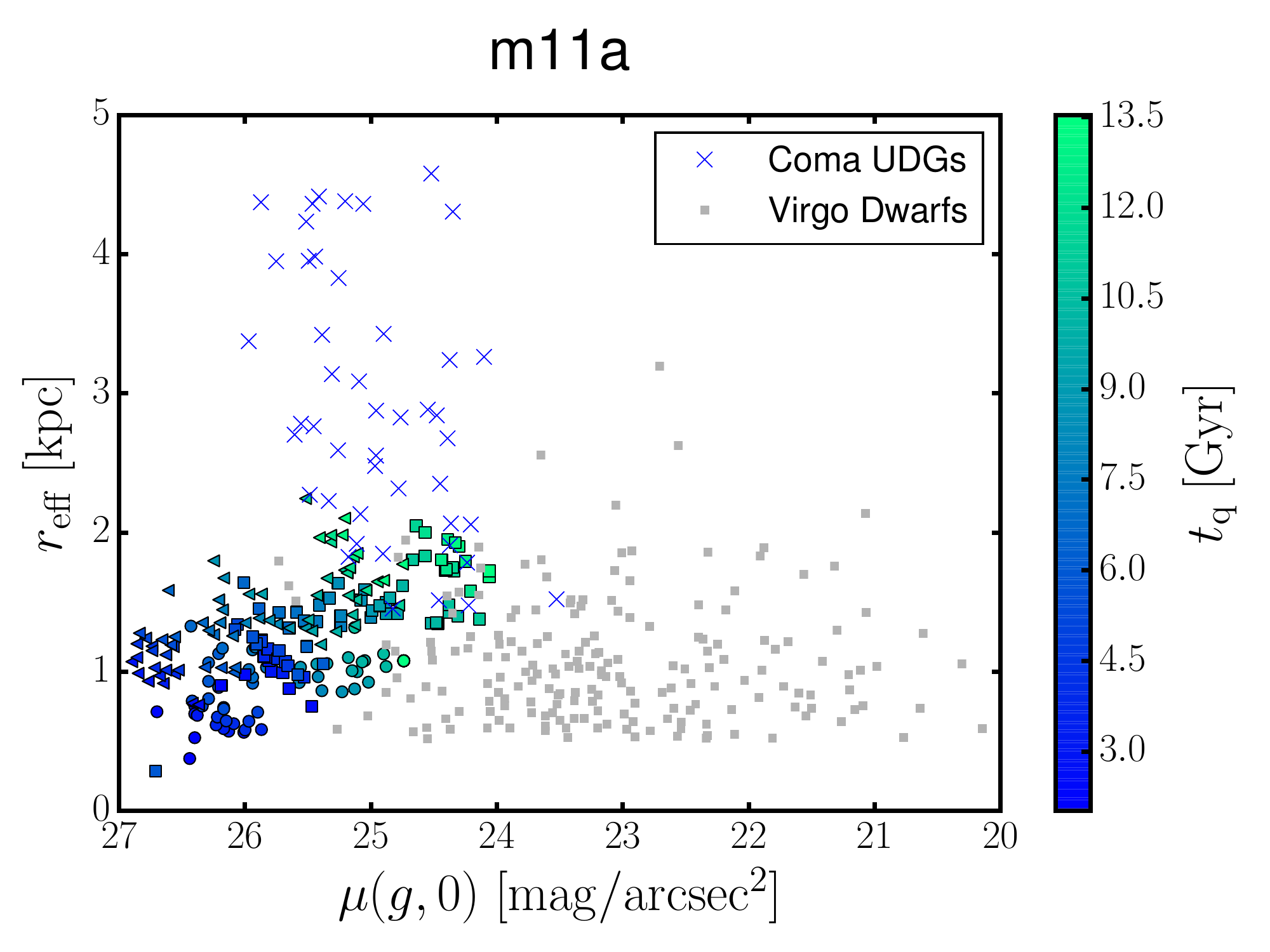}
\includegraphics[scale=0.4]{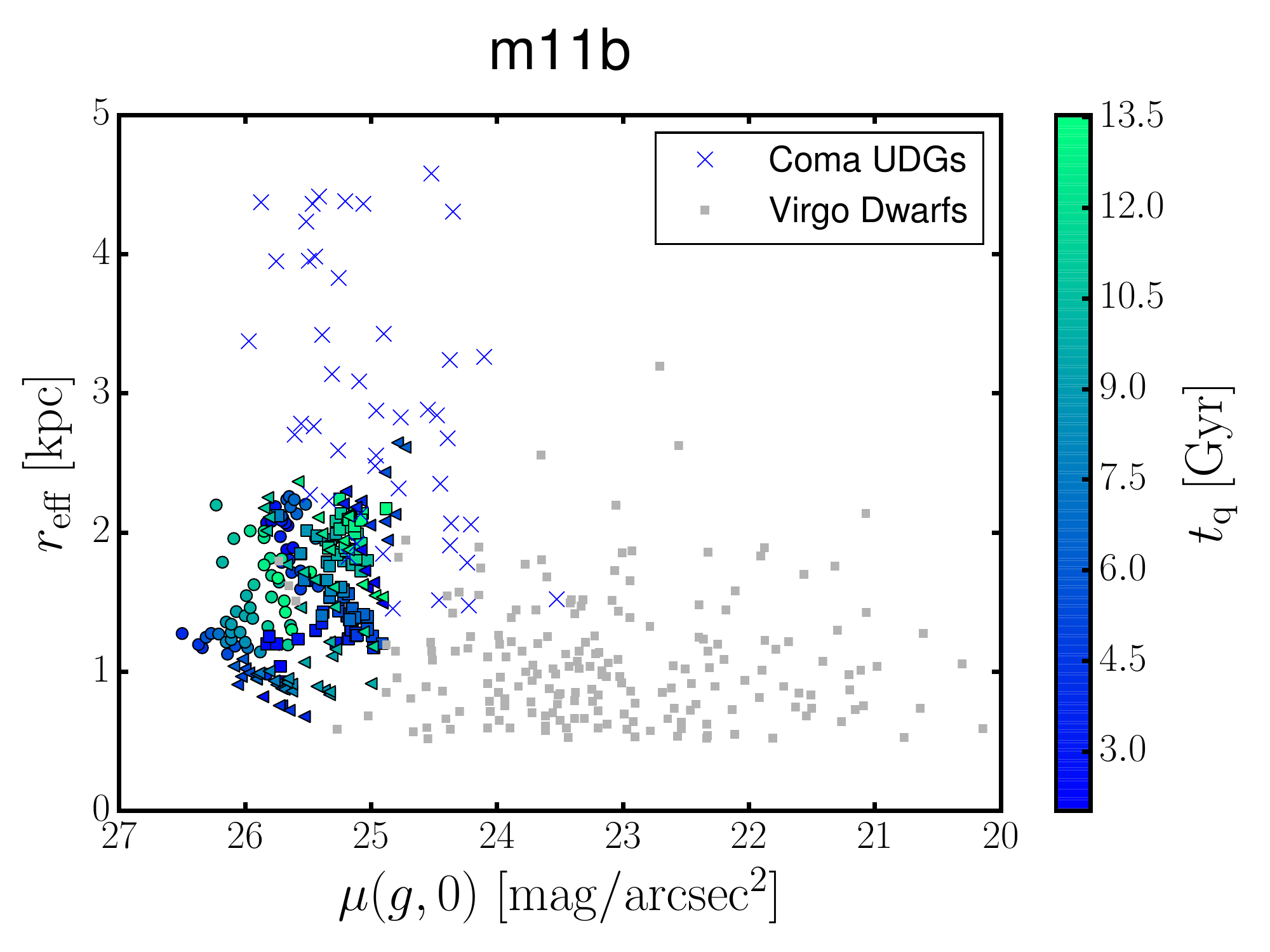}
\includegraphics[scale=0.4]{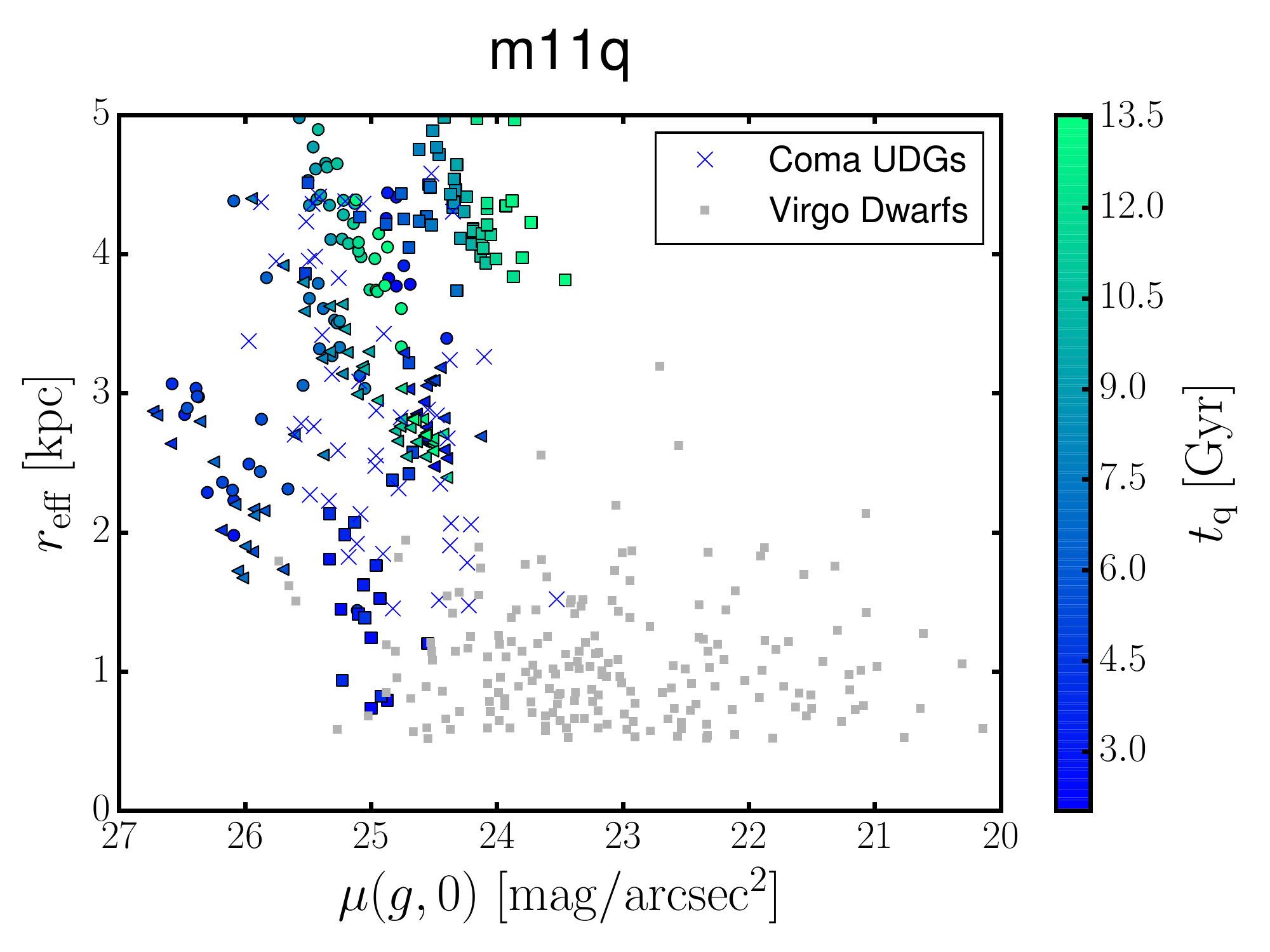}
\includegraphics[scale=0.4]{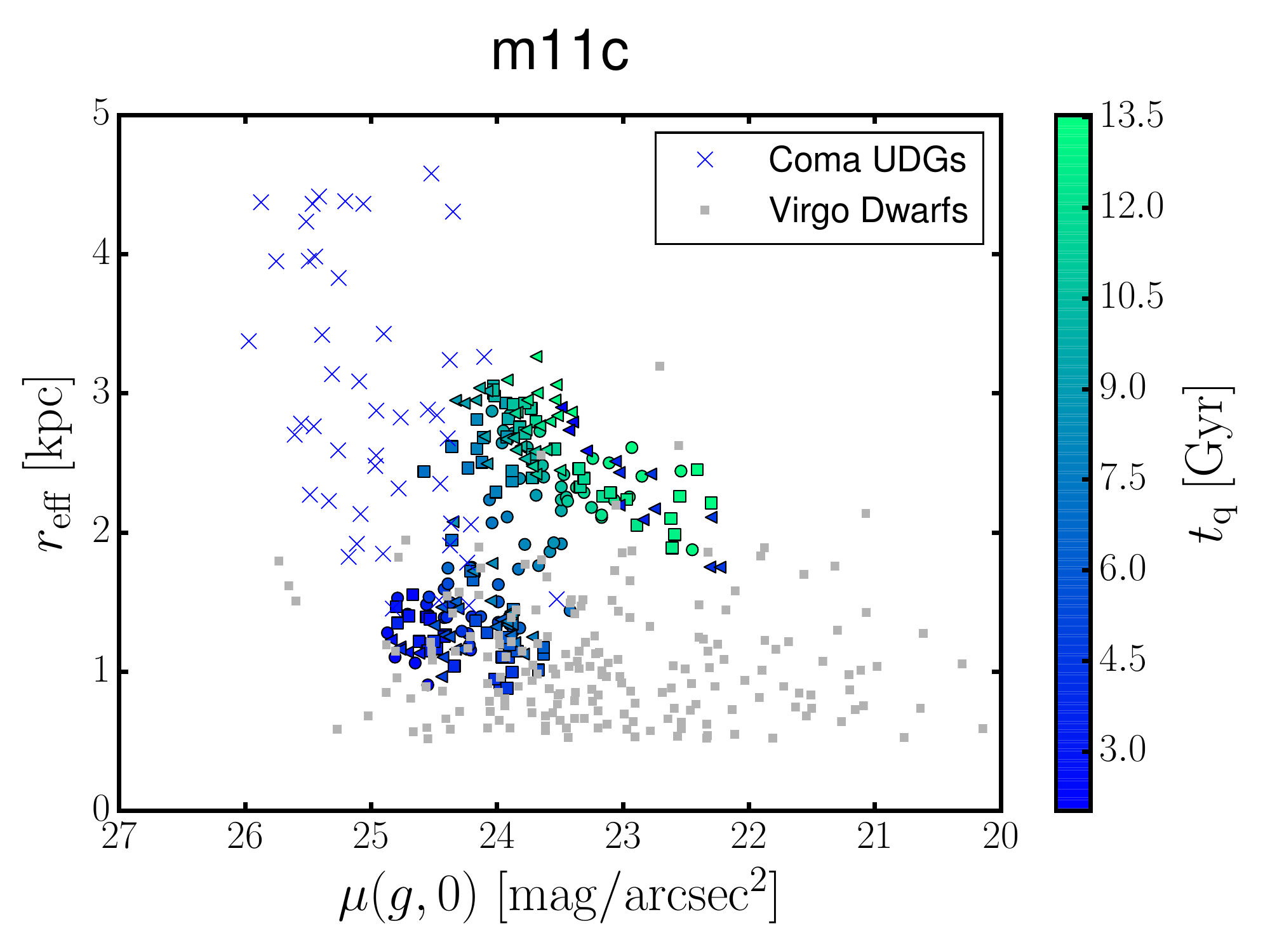}
\includegraphics[scale=0.4]{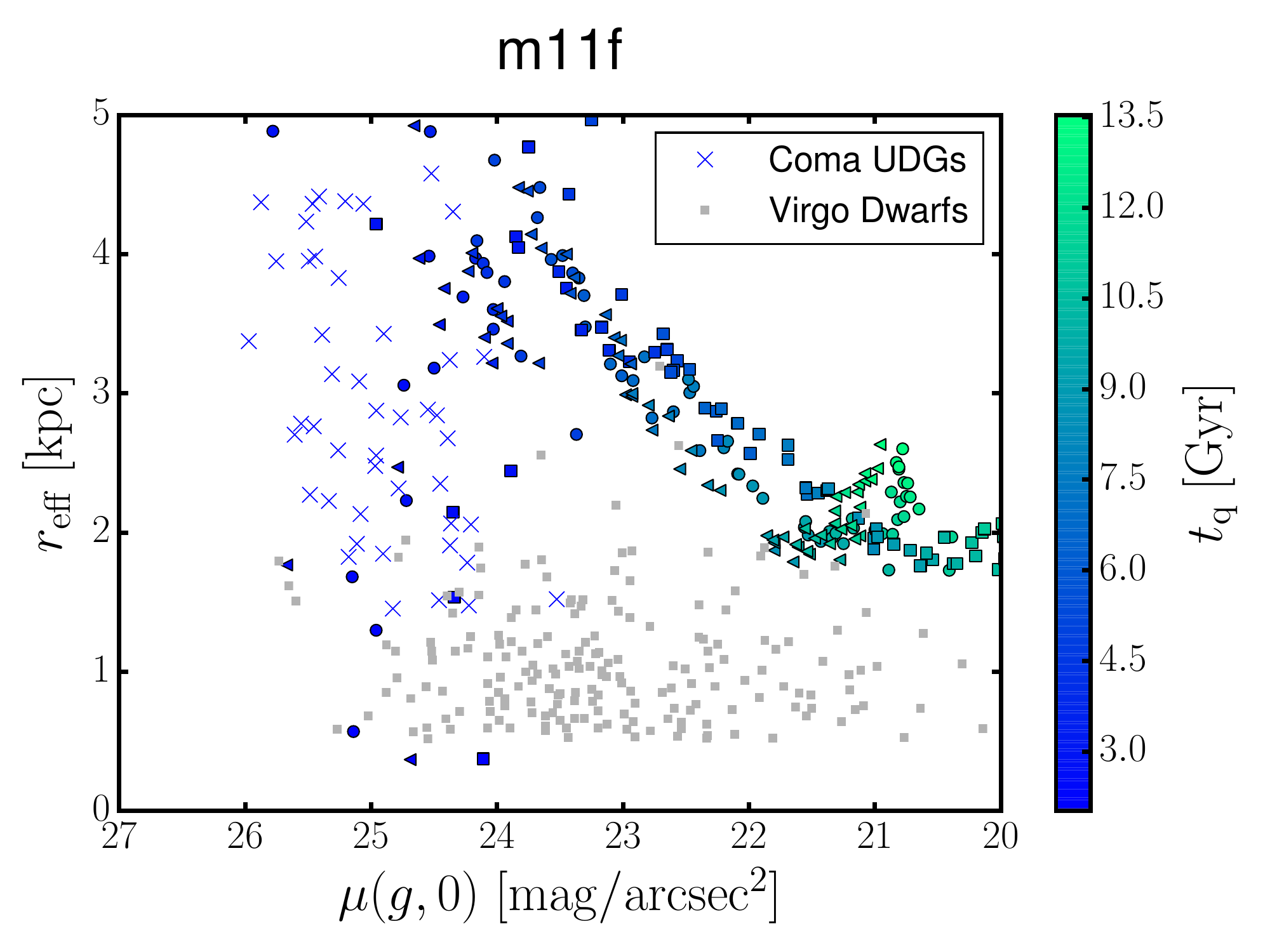}
\caption{Central {\it g}-band surface brightness of our simulated galaxies plotted against their effective radius. The colors of points represent quenching times, at which we artificially stop their star formation and passively evolve their stellar population to $z=0$ according to FSPS. The styles show different viewing angles ({\it squares}: along x axis; {\it triangles} : along y axis; {\it circles}: along z axis). Each panel represents a single simulated galaxy. We also show the observed values of early-type galaxies in the Virgo cluster \citep{Gava05} and UDGs in the Coma cluster \citep{vanD15}.  We find that all of these simulated galaxies have sizes and surface brightnesses that are consistent with observed UDGs, depending on the quenching time that we assume.}
\label{surreff}
\end{figure*}

\section{Results}
\label{results}

\subsection{Effective Radius and Surface Brightness}
\label{reffmu}

 \begin{figure*}
\includegraphics[scale=0.4]{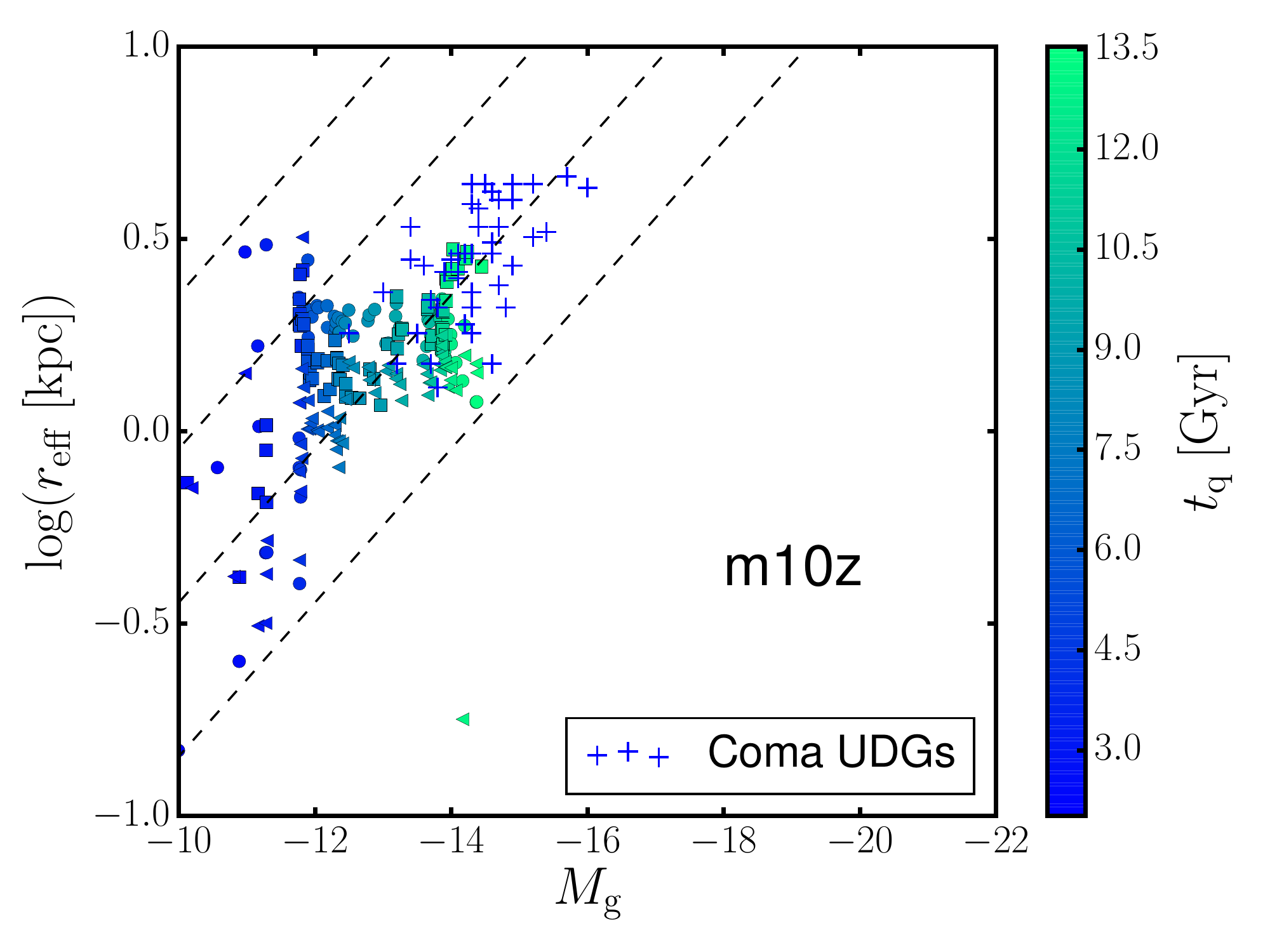}
\includegraphics[scale=0.4]{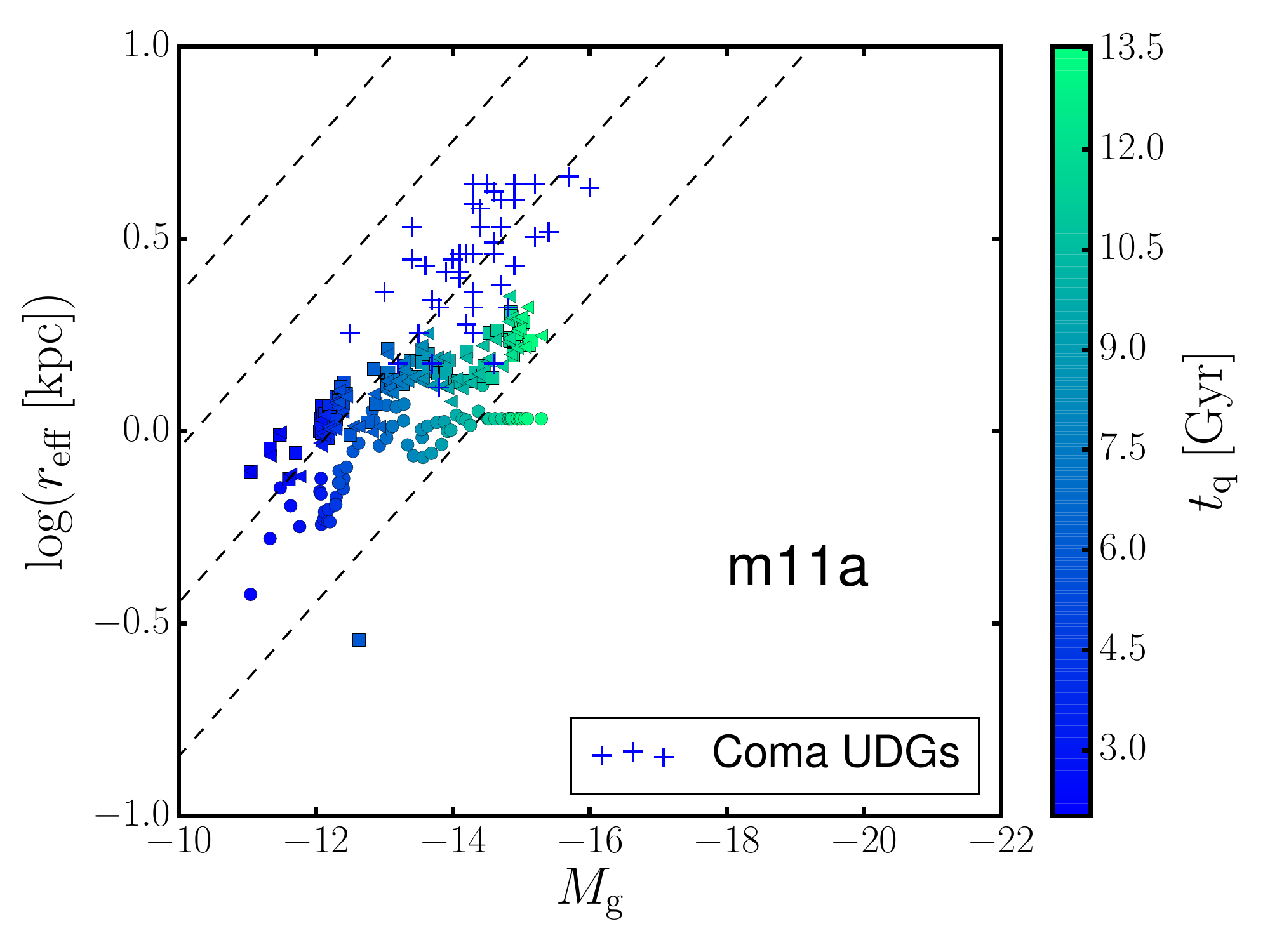}
\includegraphics[scale=0.4]{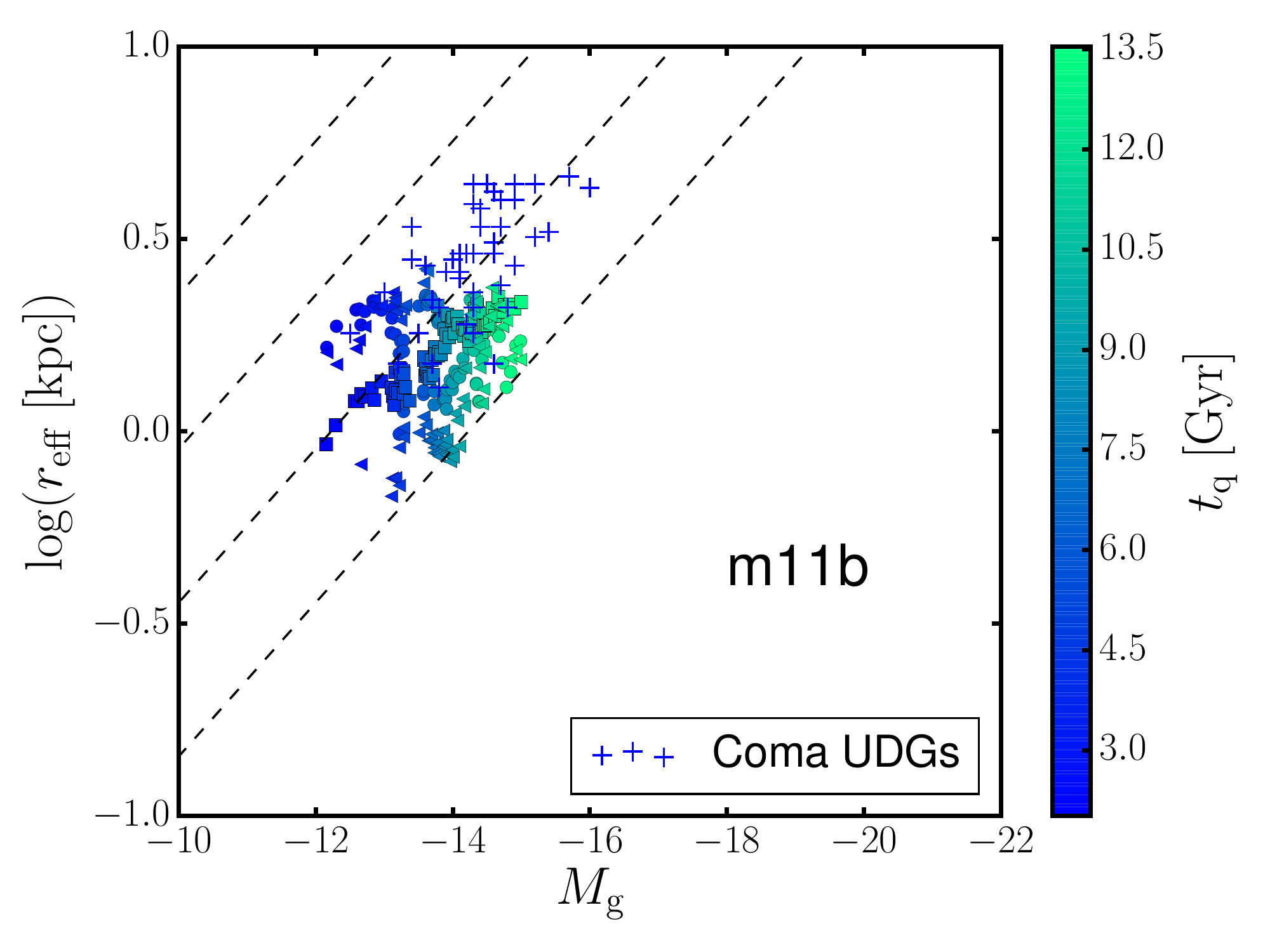}
\includegraphics[scale=0.4]{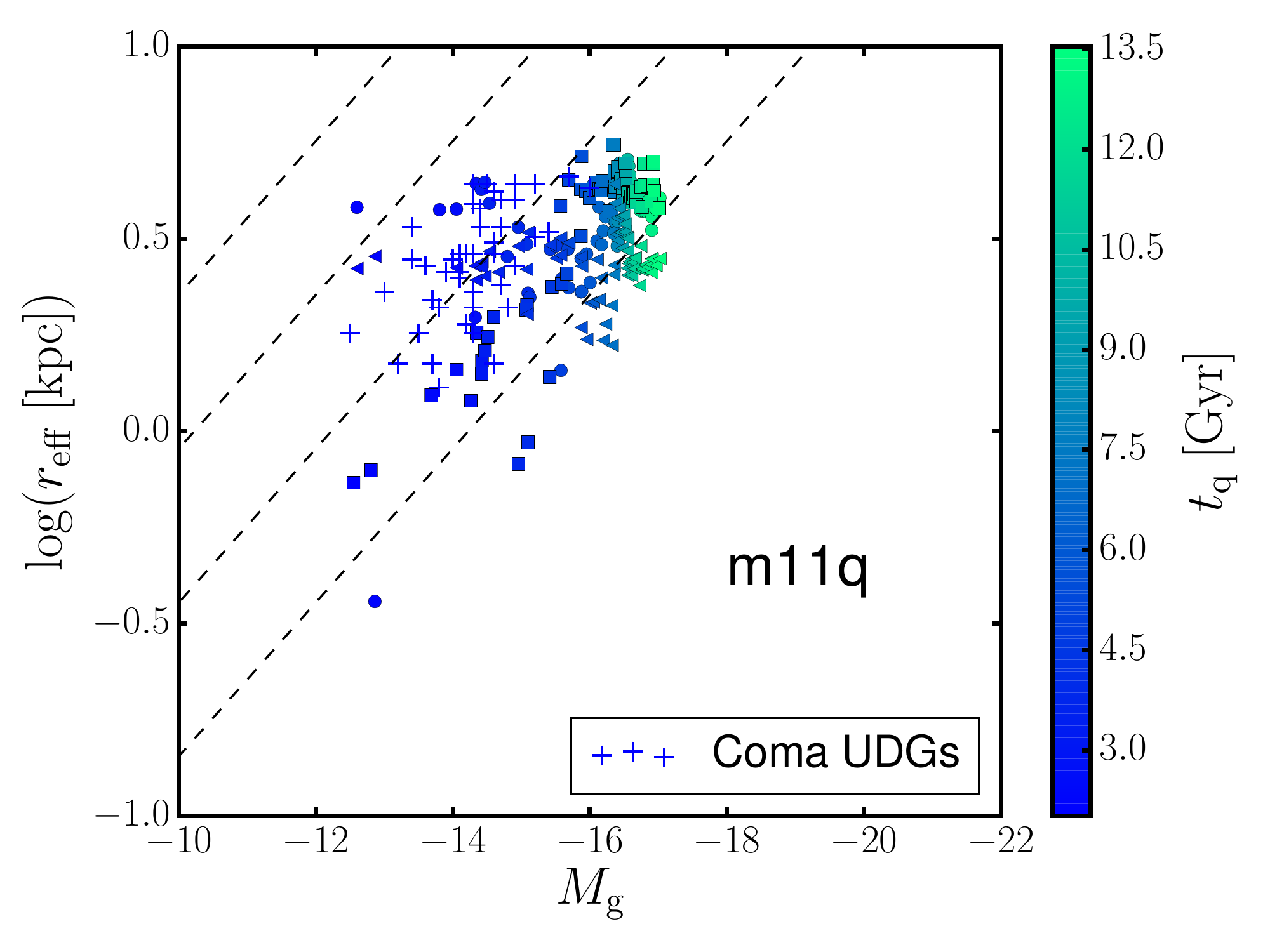}
\includegraphics[scale=0.4]{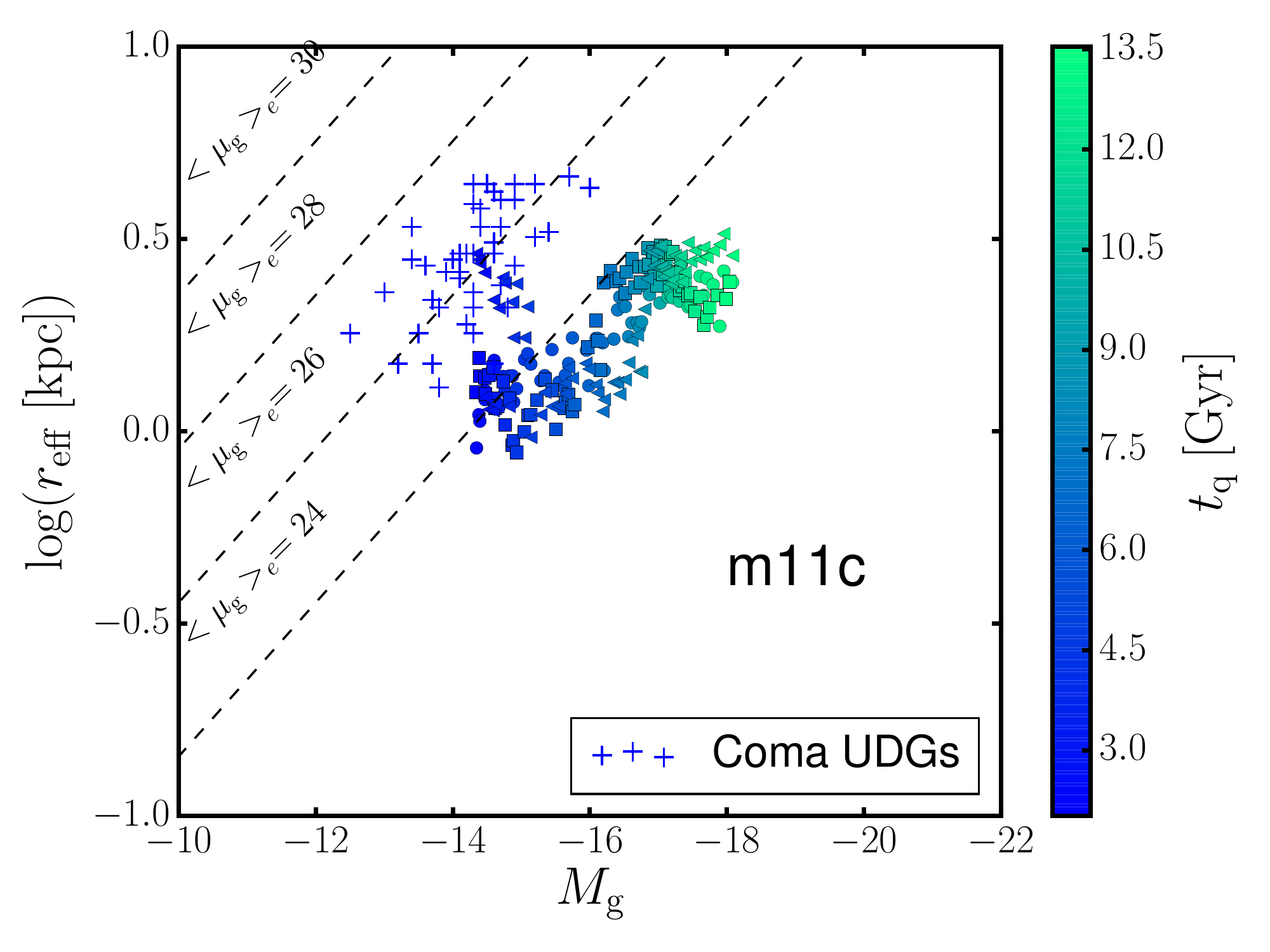}
\includegraphics[scale=0.4]{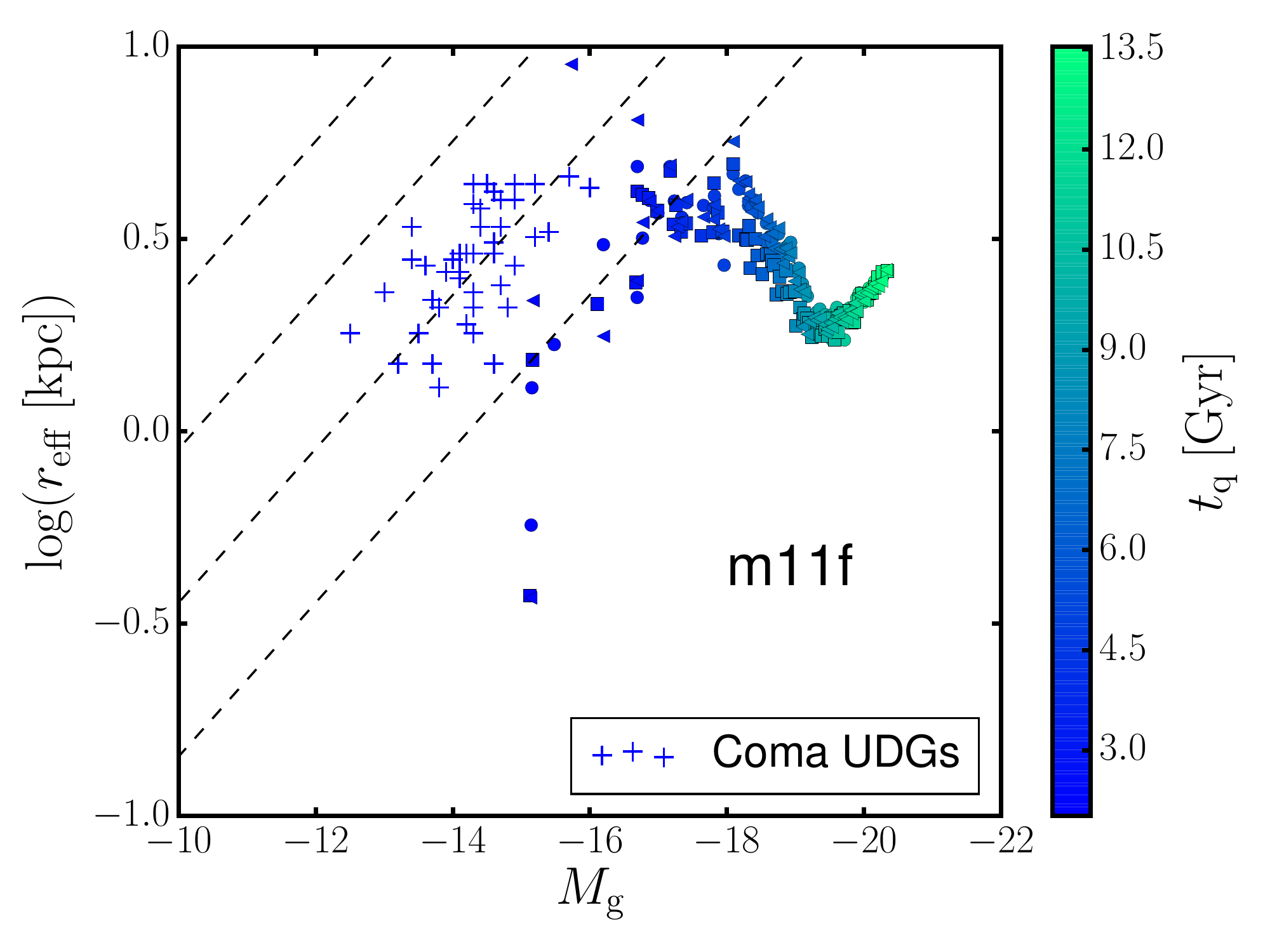}
\caption{Absolute {\it g}-band magnitude (in AB) plotted against effective radius. Dashed lines show constant {\it g}-band surface brightnesses within effective radii, $\left<\mu_g\right>_e\;[{\rm mag/arcsec^2}]$.  The color of each point represents the assumed quenching time, whereas point styles show different orientations, as in Figure~\ref{surreff}. Cross symbols represent the observed UDGs from \citet{vanD15}. Our simulated galaxies can match the range of observed surface brightness and magnitudes at some quenching times, but more massive UDGs have to quench very early to match the range of observed magnitudes.}
\label{magreff}
\end{figure*}

We first define UDGs based on the \cite{vanD15} sample: (1) $\mu(g) \gtrsim 23.5\,{\rm mag/arcsec^2}$ ; (2) $r_{\rm eff} \gtrsim 1.25 {\rm kpc}$. This definition is also similar to the selection criteria in \cite{Roma17a}. \cite{Koda15} and \citet{Miho15} also use the similar $r_{\rm eff}$ limit to define UDGs, but require different $\mu$ cutoffs according to the bands they use\footnote{Although our UDG definition does not include limits on the total magnitude, we may impose magnitude cuts to match specific observation samples in the following, e.g. Table \ref{UDGs}.}.

Figure \ref{surreff} shows the central {\it g}-band surface brightness $\mu(g,0)$, defined as the surface brightness of the fitted Sersic profile at the center, as a function of the effective radius of our galaxies, compared to the observed red UDGs and dwarf galaxies associated with galaxy clusters. We ``observe'' each galaxy along three perpendicular viewing angles for different assumed quenching times, and label the results with different symbols accordingly in Figure \ref{surreff} and \ref{magreff}.  The choice of viewing angle only mildly affects the results, consistent with expectations for roughly spheroidal geometries.  The differences in surface brightness between different viewing angles are in general smaller than 1 $\rm{mag/arcsec^2}$, except for very irregular geometries during mergers. 

In Figure \ref{surreff} we show that $\mu$ and $r_{\rm eff}$ of {\bf m10z}, {\bf m11a}, {\bf m11b}, and {\bf m11q} are a good match to the observed red UDGs for $t_{\rm q}\sim 6-13$ Gyr. Later quenching times allow these galaxies to form more stars, while strong stellar feedback increases their $r_{\rm eff}$.  {\bf m11c} and {\bf m11f} agree with the red UDGs in \cite{vanD15} only for very early quenching times ($t_{\rm q} \sim 2-4$ Gyr) but have much higher surface brightness if their star formation proceeds to later times. For early quenching times these two galaxies are therefore valid progenitors of UDGs according to our criteria stated in \S~\ref{reffmu}. If we, in addition to our standard criteria, consider the absolute magnitude range of \citet{vanD15}, {\bf m11f}, can only reproduce the bright end of the UDGs for $t_{\rm q}\lesssim 2.5$ Gyr, owing to its large g-band magnitude if quenched much later (see Figure \ref{magreff} and \ref{infallplot}). Overall, the luminosities, effective radii, and colors of stellar populations formed by $z\sim 0-3$ in dwarfs simulated with the FIRE-2 model are consistent with those of observed UDGs, but galaxies forming in more massive halos (at $z=0$) require earlier quenching times.

Figure \ref{magreff} shows effective radius, $r_{\rm eff}$, as a function of absolute {\it g}-band magnitude, $M_{\rm g}$, for the FIRE-2 dwarfs and the observed red UDGs from \cite{vanD15}. We also show lines indicating the average surface brightness within the effective radius\footnote{Note that the difference between average surface brightness within $r_{\rm eff}$ and central surface brightness, $\left<\mu_g\right>_e-\mu(g,0)$, is generally small because of relatively flat profiles, but it can occasionally reach up to $\sim 0.5 \rm mag$. }. 

All galaxies can roughly match the parameter space of the observed UDGs for a wide range of quenching times, except {\bf m11c} and {\bf m11f}, our two most massive systems. These galaxies meet our UDG criteria only for snapshots with $t_{\rm q}<3\,{\rm Gyr}$ (excluding some occasional contraction periods and minor mergers) and represent more massive UDGs. At larger $t_{\rm q}$, their surface brightnesses are higher than the observed UDGs in the \cite{vanD15} sample.

\subsection{Effects of Quenching Time}
\label{infalltime}
\begin{table*}
\small
\centering
    \begin{tabular}{llllllllllllllll}
    \hline\hline
     &\multicolumn{2}{c}{Observed} &\multicolumn{2}{c}{\bf m10z} &\multicolumn{2}{c}{\bf m11a} &\multicolumn{2}{c}{\bf m11b}&\multicolumn{2}{c}{\bf m11q} &\multicolumn{2}{c}{\bf m11c} & \multicolumn{2}{c}{\bf m11f}\\
    \hline

\rule{0pt}{3ex} $t_{\rm q} [{\rm Gyr}]$                   &- &      &$13.4$& $\,_{8.2}^{13.5} $&      $10.4$& $\,_{5.7}^{13.5} $&      $10.7$& $\,_{2.6}^{13.5} $&      $2.7$& $\,_{2.3}^{6.0} $&      $2.0$& $\,_{2.0}^{6.5} $&      $2.2$& $\,_{2.2}^{2.4} $\\ [1.5mm]

$r_{\rm eff} [{\rm kpc}]$                      &2.8&$\,_{1.5}^{4.6}$  &$2.9$& $\,_{1.4}^{3.2} $&      $1.4$& $\,_{1.3}^{1.4} $&      $2.0$& $\,_{1.3}^{2.0} $&      $1.7$& $\,_{1.3}^{4.0} $&      $1.3$& $\,_{1.3}^{1.5} $&      $1.6$& $\,_{1.6}^{2.5} $\\ [1.5mm]

$\mu({\rm g},0)$                                  &25.0&$\,_{23.5}^{26.5}$ & $25.67$& $\,_{26.3}^{25.67} $&      $24.42$& $\,_{25.96}^{23.76} $&      $25.14$& $\,_{25.77}^{24.92} $&      $25.98$& $\,_{26.01}^{25.47} $&      $23.74$& $\,_{23.74}^{23.4} $&      $26.24$& $\,_{26.24}^{24.87} $\\ [1.5mm]

${ M}_{\rm g}  [{\rm AB}]$                       &-14.3&$\,_{-12.5}^{-16.0}$ &$-14.3$& $\,_{-12.5}^{-14.4} $&      $-14.3$& $\,_{-12.5}^{-15.2} $&      $-14.3$& $\,_{-12.7}^{-15.0} $&      $-14.3$& $\,_{-13.8}^{-16.0} $&      $-14.3$& $\,_{-14.3}^{-16.0} $&      $-15.2$& $\,_{-15.2}^{-16.0} $\\ [1.5mm]

$g-i$                                                     &0.8&$\,_{0.9}^{0.7}$  & $0.54$& $\,_{0.77}^{0.51} $&      $0.72$& $\,_{0.79}^{0.59} $&      $0.76$& $\,_{0.79}^{0.59} $&      $0.81$& $\,_{0.8}^{0.85} $&      $0.84$& $\,_{0.84}^{0.86} $&      $0.83$& $\,_{0.83}^{0.83} $\\ [1.5mm]



${\rm Age_* [ Gyr]}$                            &-   &     &$5.4$& $\,_{9.3}^{5.4} $&      $6.8$& $\,_{11.0}^{5.1} $&      $8.7$& $\,_{12.2}^{7.2} $&      $11.6$& $\,_{11.8}^{9.9} $&      $12.6$& $\,_{12.6}^{9.9} $&      $12.1$& $\,_{12.1}^{11.9} $\\ [1.5mm]

${\rm [Fe/H]}$                                     &-   &     & $-1.41$& $\,_{-1.67}^{-1.41} $&      $-1.26$& $\,_{-1.7}^{-1.1} $&      $-1.28$& $\,_{-1.77}^{-1.17} $&      $-1.54$& $\,_{-1.71}^{-1.15} $&      $-1.39$& $\,_{-1.39}^{-1.02} $&      $-1.45$& $\,_{-1.45}^{-1.37} $\\ [1.5mm]

$M_*[10^8\msun]$                              &-   & &$0.53$& $\,_{0.19}^{0.55} $&      $0.8$& $\,_{0.21}^{1.22} $&      $0.94$& $\,_{0.26}^{1.16} $&      $1.2$& $\,_{0.74}^{5.58} $&      $1.36$& $\,_{1.36}^{5.76} $&      $2.77$& $\,_{2.77}^{6.12} $\\ [1.5mm]    

$M_{\rm h,q} [10^{10}\msun]$                             &- & &$3.5$& $\,_{3.4}^{3.5} $&      $3.8$& $\,_{2.9}^{4.1} $&      $4.0$& $\,_{1.3}^{4.4} $&      $3.4$& $\,_{3.1}^{9.3} $&      $2.2$& $\,_{2.2}^{11.4} $&      $5.5$& $\,_{5.5}^{9.7} $\\ [1.5mm]

$10^3M_*/M_{\rm h,q}$                             &- & &$1.53$& $\,_{0.55}^{1.59} $&      $2.04$& $\,_{0.72}^{2.94} $&      $2.36$& $\,_{1.92}^{2.65} $&      $3.75$& $\,_{1.71}^{6.09} $&      $6.14$& $\,_{6.14}^{4.93} $&      $4.91$& $\,_{4.91}^{5.94} $\\ [1.5mm]

$M_{\rm 1/2,obs}[10^8\msun]$                             &-  &&      $4.28$& $\,_{2.29}^{4.98} $&      $3.41$& $\,_{2.16}^{2.5} $&      $7.65$& $\,_{2.6}^{9.3} $&      $2.1$& $\,_{1.55}^{21.54} $&      $3.68$& $\,_{3.68}^{3.0} $&      $13.66$& $\,_{13.66}^{14.25} $\\ [1.5mm]

$M_{\rm 1/2}[10^8\msun]$                             &-  &&      $8.83$& $\,_{4.67}^{9.53} $&      $4.76$& $\,_{2.82}^{4.66} $&      $7.68$& $\,_{4.87}^{9.92} $&      $20.48$& $\,_{52.56}^{32.83} $&      $10.06$& $\,_{10.06}^{30.13} $&      $25.1$& $\,_{25.1}^{26.0} $\\ [1.5mm]

$10 \times f_{\rm 1/2,*}$                             &- & &      $0.39$& $\,_{0.26}^{0.4} $&      $0.86$& $\,_{0.37}^{1.2} $&      $0.6$& $\,_{0.32}^{0.58} $&      $0.27$& $\,_{0.05}^{0.99} $&      $0.91$& $\,_{0.91}^{0.73} $&      $0.64$& $\,_{0.64}^{0.83} $\\ [1.5mm]
    \hline
    \end{tabular}
   \caption{Characteristic properties of simulated UDGs. 
   The properties of the simulated galaxies are extracted for quenching times when they satisfy UDG selection criteria from \S~\ref{reffmu} and when, in addition, their ${M}_{\rm g}$ falls within the observed range of red UDGs in \citet{vanD15}, ${M}_{\rm g}$=[-16.0, -12.5]. The values presented in large numbers are determined at the $t_{\rm q}$ for which the g-band magnitude is the closest to the median observed magnitude $M_g=-14.3$, while the small numbers show the maximum and minimum during the range of $t_q$ described above. Effective radius, $r_{\rm eff}$ and central surface brightness, $\mu(g,0)$ are determined by GALFIT. The absolute magnitude, $M_g$ and color, $g-i$, are determined directly from star particles. Stellar age and metallicity are mass weighted. Stellar mass is measured within 0.2$R_{\rm vir}$. Next, we show the range of halo masses, $M_{\rm h,q}$, and the stellar-to-halo mass ratio, $M_*/ M_{\rm h,q}$. $M_{\rm 1/2,obs}$ is the total stellar plus DM mass within de-projected half-light radius ($r_{\rm 1/2}=r_{\rm eff} \times \sqrt{b/a}\times 4/3$, where b/a is axis ratio from GALFIT), whereas $M_{\rm 1/2}$ is the total stellar plus DM mass within 3D half stellar mass radius. In addition, we show the ratio of the stellar mass to stellar plus DM mass, $f_{\rm 1/2, *}$, within the de-projected half light radius $r_{\rm 1/2}$. All of the masses and mass ratios are measured at $t_{\rm q}$ (i.e. we assume that structural properties of galaxies and halos remain fixed after $t_{\rm q}$).  The second column ("Observed") shows properties of observed UDGs. Given the constraints on effective radius, surface brightness and g-band luminosity/magnitude, we predict colors, ages, metallicities and stellar-to-halo mass ratios. All quantities are measured as viewed along x axis.
}
    \label{UDGs}
\end{table*}
 \begin{figure*}
\includegraphics[scale=0.35]{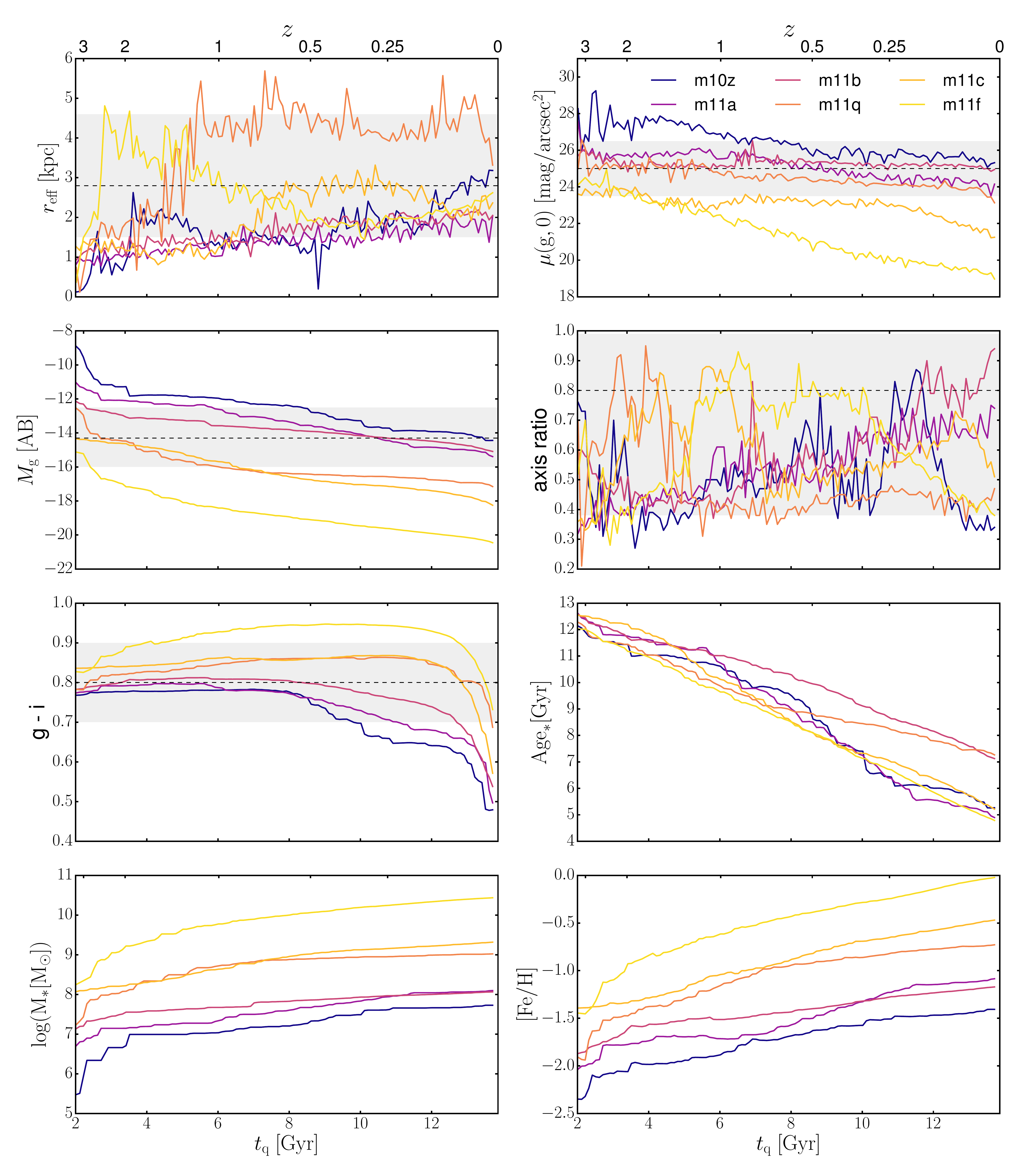}
\caption{Effective radius $r_{\rm eff}$, central {\it g}-band surface brightness $\mu(g,0)$, absolute {\it g}-band magnitude $M_{\rm g}$, axis ratio, g - i color, (mass-weighed) stellar age ${\rm Age_*}$, stellar mass within $0.2 R_{\rm vir}$, $M_*$, as functions of quenching time. We use GALFIT to determine central surface brightness from the {\it g}-band images and effective radii and axis ratios from g + i images. All other quantities are obtained directly from star particles without using a fit. All quantities are measured as viewed along x axis. Horizontal dashed lines show the median values of the observed UDGs \citep{vanD15} while horizontal shaded regions show their ranges.}
\label{infallplot}
\end{figure*}
In this section, we explore how properties of our simulated galaxies at $z=0$ depend on quenching time. We explore the range of the allowed quenching times by matching the observed properties of our simulated UDGs, e.g. $r_{\rm eff}$, $M_{\rm g}$ and $\mu_{\rm g}$, to the observations.

We plot the properties of our galaxies as a function of quenching time in Figure \ref{infallplot} and compare them with the median values from the observations \citep{vanD15}, shown with horizontal dashed lines. The gray regions show the minimum and maximum observed values for $r_{\rm eff}$, $M_{\rm g}$ and $\mu_{\rm g}$. We also show the evolution of axis ratio, g-i color, stellar mass $M_{*}$, stellar age $\rm{Age}_*$, and metallicity ${\rm [Fe/H]}$.

Absolute {\it g}-band magnitudes and mass weighted stellar ages drop with quenching time for all galaxies because later quenching implies a shorter passive evolution period and a longer time during which a galaxy can form stars. The typical axis ratios of our simulated galaxies are $\sim$ 0.8 since our galaxies are usually spheroidal owing to continuous stellar feedback, which prevents formations of prominent disks in dwarfs \citep{Whee17,ElBa17b}. Our most massive galaxy builds a stellar disk at late times, which shows as a fast drop in axis ratio.

Finally, the g-i colors of all but our most massive dwarfs are approximately $0.75 - 0.85$ for $t_{\rm q} \lesssim 10-11\,{\rm Gyr}$, consistent with observations of UDGs. The slow change in colors is caused by the interplay between increasing metallicities and decreasing mean stellar ages as we increase $t_{\rm q}$. These competing effects mostly cancel out and prevent strong changes in the overall colors of the galaxies until $t_{\rm q}\sim 10\,{\rm Gyr}$ for lower mass dwarfs and $t_{\rm q}\sim 12\,{\rm Gyr}$ for our higher mass dwarfs. This implies that mean stellar ages and metallicities of UDGs cannot be determined with g-i colors alone. 

Although we cannot infer precise quenching times of observed UDGs from their g-i colors only, effective radius, surface brightness, and {\it g}-band magnitude can provide tighter constraints. {\bf m10z}, {\bf m11a} and {\bf m11b} correspond to red UDGs if $t_{\rm q}\gtrsim 5\,{\rm Gyr}$.  The more massive dwarfs {\bf m11q} and {\bf m11c} must have $t_{\rm q}\lesssim  6\,{\rm Gyr}$ and $t_{\rm q}\lesssim  3\,{\rm Gyr}$, respectively, in order to match the {\it g}-band magnitude of the sample in \citet{vanD15}. 

Stellar age and metallicity measurements also give useful constraints on quenching time. \cite{Kado17} found low stellar metallicity (${\rm[Fe/H]}\lesssim-1.5$) from the stacked spectrum of Coma UDGs. \cite{Gu17} analyzed the optical spectra of three of the brightest Coma UDGs and found they are metal poor (${\rm[Fe/H]}=-0.8^{+0.5}_{-0.5}-1.3^{+0.4}_{-0.4}$) and old ($7.9^{+3.6}_{-2.5}-9.1^{+3.9}_{-5.5}$ Gyr). Using optical through near infrared SED fitting, \cite{Pand17} inferred a UDG in the Virgo cluster (VCC 1287) to be metal poor (${\rm[Fe/H]}\lesssim-1.0$) and old ($\gtrsim 9$ Gyr). These data and Figure \ref{infallplot} together imply an early quenching scenario for cluster UDGs ($t_{\rm q}\lesssim  6\,{\rm Gyr}$).

\subsection{Characteristic properties of simulated red UDGs}

Table \ref{UDGs} lists the properties of our simulated galaxies, for a range of quenching times during which they match the observed range of $r_{\rm eff}$ and $\mu(g,0)$. We also list the median values of the observed UDGs \citep{vanD15}. When the {\it g}-band magnitude of our simulated galaxy matches the median of the observed sample at a certain quenching time, the galaxy also yields a close match in effective radius, central surface brightness, and color, suggesting that our simulated galaxies are good analogs of the observed UDGs. While our galaxies are slightly less spheroidal than UDGs, we expect they would be rounder if the dynamical effect of gas removal were taken into account (shown in Appendix \ref{agasre}). Simulated and observed UDGs have similar stellar masses, largely determined by the absolute magnitude selection of the sample, since old stars have approximately constant stellar mass-to-light ratios. 

With earlier quenching times, stars are older at $z = 0$ and have lower metallicities because of the evolution of the galaxy mass-metallicity relation \citep[e.g.][]{Zahi13,Ma16}. Metallicities of UDGs could potentially be used to constrain quenching times of UDGs. However, for our simulated sample, typical metallicities grow very slowly with later quenching times because the metallicity evolution is offset by larger stellar masses of the galaxies that satisfy the UDG selection for early quenching times (typically galaxies with $z=0$ halo mass $\gtrsim 10^{11}\msun$). A larger sample of simulated galaxies is needed to explore the metallicity trends in detail. 

Observationally, long exposure spectroscopic studies combined with stellar population modeling are necessary to determine dynamical masses,  stellar ages and metallicities of UDGs and potentially constrain their origin and quenching times \citep[e.g.][]{Maka15,Kado17,Gu17,Pand17}. Low surface brightness sensitive instruments such as the Keck Cosmic Web Imager \citep{Mart10} should be helpful in extending such studies to a larger number of objects.  Overall, for the observed range of surface brightnesses and magnitudes of UDGs, our model predicts a uniform population of galaxies in terms of their stellar masses and g-i colors but with a broad range of average stellar ages.

\subsection{UDGs and their halos}

\begin{figure}
\includegraphics[scale=0.35]{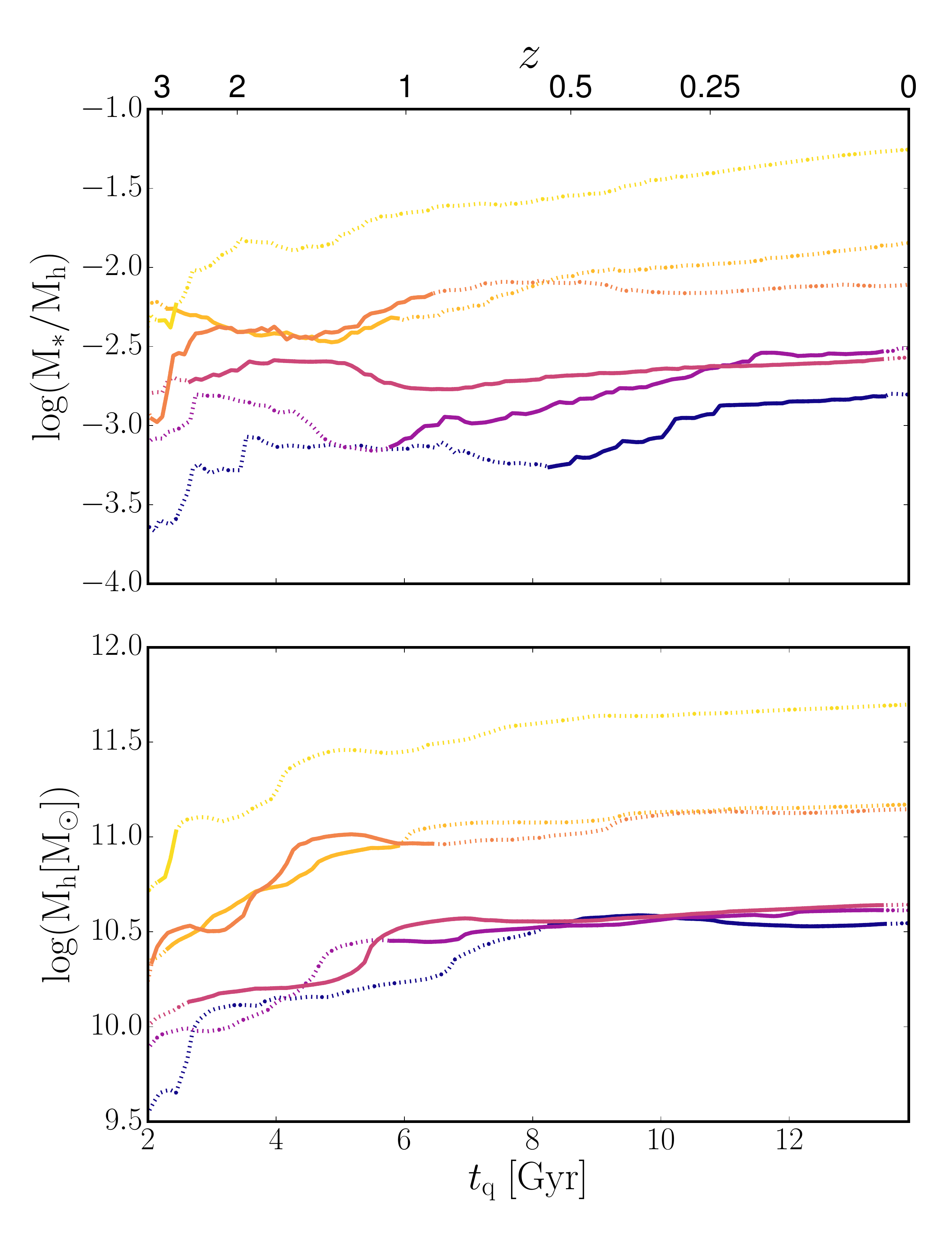}
\caption{({\it Upper}) stellar fraction and ({\it lower}) halo mass as functions of quenching time for our sample of simulated galaxies. Solid lines show the quenching times when they satisfy the UDG criteria and dotted lines when they do not. The color scheme follows Figure \ref{infallplot}. Our UDGs have ``normal'' stellar fraction for they stellar and halo mass range ($\sim 10^{-2}-10^{-3}$).}
\label{Msh}
\end{figure}

The $M_*/M_{\rm h}$ row in Table \ref{UDGs} shows the stellar-to-halo mass ratios measured at the range of quenching times when these galaxies satisfy red UDG criteria.  The upper panel of Figure \ref{Msh} shows the ratios as a function of quenching time. While the ratios are small, they do not deviate much from the dwarfs with similar masses in the Local Group (LG, \citealt{McCo12}). It is reasonable to expect that DM halos of UDGs cannot grow significantly after $t_{\rm q}$ due to cluster influence. Furthermore, the outskirts of UDG halos will likely be stripped or modified following the infall into a cluster. Stellar to total mass (stellar+DM) ratio within effective radius, given in the last row of Table \ref{UDGs}, is thus a more robust quantity.  This ratio shows that for quenching times when our galaxies are analogs to UDGs, their central regions are strongly DM dominated.

Recently, \citet{Beas16a} claimed that one of the UDGs in the Virgo cluster, VCC 1287, has a stellar fraction $\sim 3.5\times 10^{-4}$, much lower than ``normal'' dwarf galaxies of similar masses whose stellar fractions are $\sim 10^{-2}-10^{-3}$. They estimated its total halo mass using the velocity dispersion and the number of GCs. While at face value this is lower than the stellar fractions of our simulated dwarfs, they do not directly measure the mass at the virial radius; instead, they infer it assuming a density profile, which introduces significant uncertainty.  However, their observations do directly constrain the mass within 8.1 kpc with the ``trace mass estimator'' \citep{Watk10} using GCs as tracers; we therefore compare their estimate to the enclosed masses of our simulated galaxies measured at this radius.

\begin{figure}
 \includegraphics[scale=0.45]{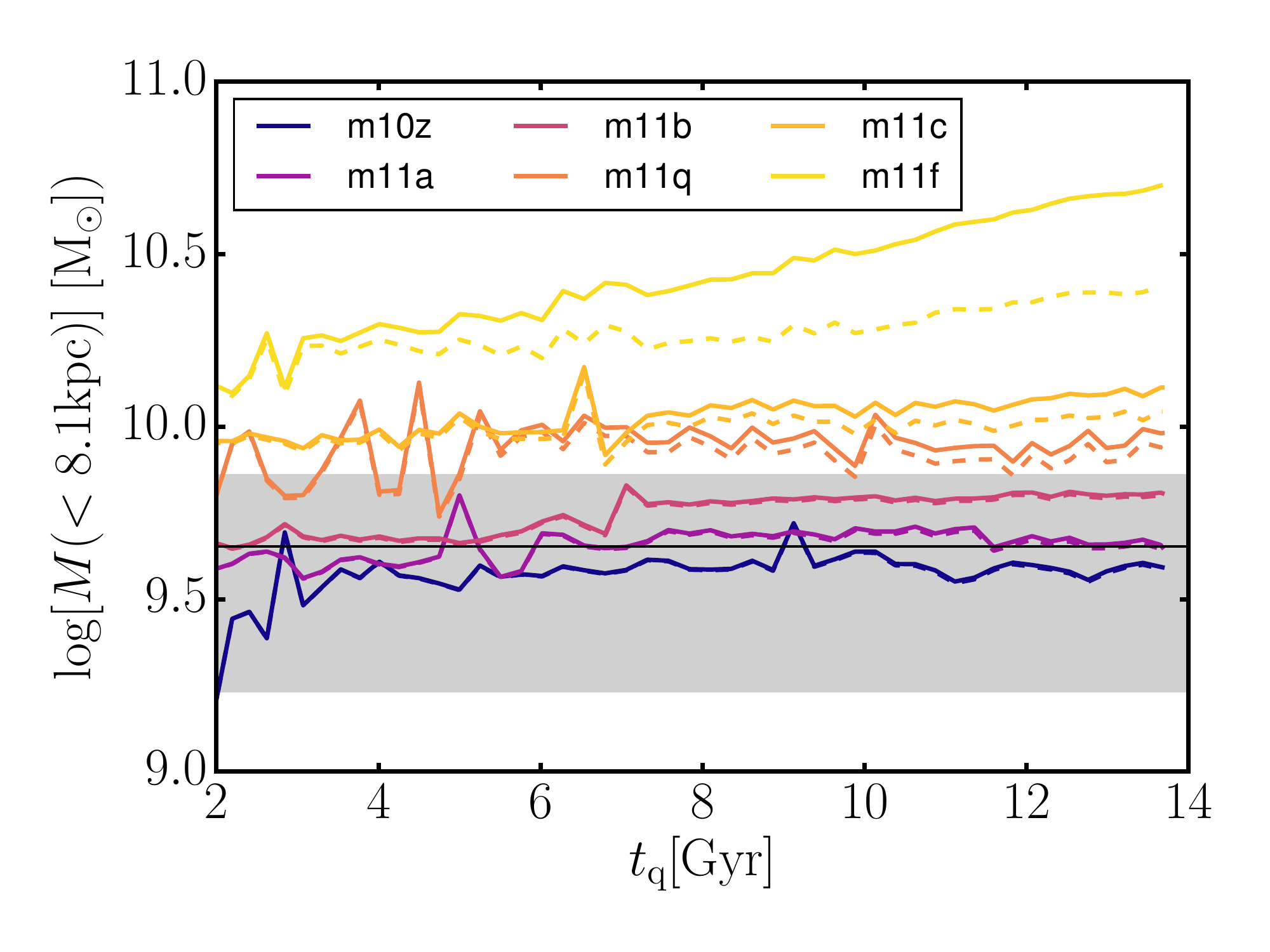}
\caption{Enclosed stellar plus DM (solid) and DM only (dashed) masses of our simulated halos within 8.1 kpc as a function of $t_{\rm q}$ compared to VCC 1287 \citep{Beas16a} (black horizontal line; shaded region indicates the uncertainties in the observation). The colors follow Figure \ref{infallplot}. The enclosed masses of {\bf m10z}, {\bf m11a} and {\bf m11b} roughly match the observed value for a broad range of $t_{\rm q}$.}
\label{Menc}
\end{figure}

Figure \ref{Menc} shows the enclosed stellar and DM mass within 8.1 kpc at different $t_{\rm q}$. We do not include the gas mass as we are comparing to a red cluster UDG that has likely lost its interstellar gas. The enclosed masses of our simulated dwarfs are roughly constant over 10 Gyr, and our lower mass halos {\bf m10z, m11a and m11b} can match the measured enclosed mass in VCC 1287, while having normal stellar fractions. Therefore, our simulated red UDGs formed as regular dwarf galaxies. However, their growths stopped at early times while outskirt of isolated halos can continue growing until much later. We may wrongly infer much larger masses of their host halos from the measurements of the central regions of UDGs and the mass profiles of isolated halos.

The enclosed masses within the inner 8.1~kpc of several of our dwarfs match the value for VCC 1287 for long periods of time, during which their total halo masses change significantly. For example, the halo mass of {\bf m10z} grew by a factor of six (see Figure \ref{Msh}) while the mass within 8.1~kpc remained consistent with VCC 1287. This constancy (lack of growth) of the inner DM profile is typical in $\Lambda$CDM: most “growth” in low-mass halos at late cosmic times occurs because of a drop in the reference density, not because of a change in the mass enclosed within a fixed physical radius (“pseudo-evolution”, \citealt{Diem13,vand14,Wetz15}). The exception is {\bf m11f}, which forms the most massive halo in our sample and whose stellar mass grew rapidly at late times, contributing a significant fraction of mass within 8.1~kpc  and contracting the underlying DM profile \citep[see e.g.][]{Chan15}, although the overall effect is small as shown in Figure \ref{Menc}. 

Given the board range of possible quenching times and the constancy of inner halo mass, one cannot accurately estimate their total halo mass and stellar-to-halo mass ratios from their enclosed masses without knowing their quenching times. If we determine their total halo mass by comparing their inner mass and the $z=0$ halo mass profile of an isolated halo, the estimated halo mass may be several times larger than the true value.

\section{Discussion}
\label{discussion}

\subsection{UDG mass estimates}
\label{progenitor}
While recent observations have constrained the masses of the inner halos of several UDGs, the fraction of known UDGs with accurate measurements for the inner halo mass is very small. For those few systems with their inner masses measured, moreover, one must assume a density profile and extrapolate to infer a total halo mass.  Together, these complications have led to two different views for the characteristic mass of UDG host halos. 

\citet{vanD15,vanD16} suggested that UDGs are ``failed'' ${\rm L}\star$ galaxies, whereas \cite{Yozi15}, \cite{Amor16}, \cite{Beas16a}, \cite{Beas16b} and \cite{Peng16} argued they are ``failed'' {\it dwarf} galaxies, based on the inferred ratios of stellar-to-halo mass. Both camps support their claims with the enclosed masses of UDGs inferred from velocity dispersions and the numbers of globular clusters (GCs). 

In the previous section, we showed that simulated UDGs can form in halos with $M_{\rm h}\sim 3-15 \times 10^{10}\msun$ and that their central enclosed masses alone are not be a good indicator of their host halo mass at the time of quenching. Here we compare the velocity dispersions and masses of our simulated UDGs to the several observed examples and critically examine the methodology used to infer masses from observations. 

In Table \ref{veldis} we show the range of velocity dispersion as seen along three perpendicular directions 
and the average values of each of our UDG analogs. The line-of-sight velocity dispersion of our galaxies ranges from $\sim 20-50$km/s. Our intermediate mass dwarfs provide a good match to $\left \langle \sigma \right \rangle=33^{+16}_{-10}$km/s measured for VCC 1287 \citep{Beas16a} and our most massive UDG provides a match to $\left \langle \sigma \right \rangle=47^{+8}_{-6}$km/s measured for Dragonfly 44 \citep{vanD16}.

With the measured velocity dispersion and effective radius, \citet{vanD16,Beas16a} and \citet{Beas16b} inferred the enclosed mass within stellar half-light radius using an equation first presented in \citet{Wolf10}:
\begin{equation} 
M_{\rm 1/2}\simeq 9.3\times10^5\left ( \frac{\left<\sigma^2_{\rm los}\right>}{\rm km^2/s^2} \right )\left ( \frac{r_{\rm eff}}{\rm kpc} \right )\msun,
\label{Wolf}
\end{equation}
where $M_{\rm 1/2}$ is the total mass within the 3D half light radius and $\left<\sigma^2_{\rm los}\right>$ is the square of the line-of-sight velocity dispersion. Strictly speaking this relation is only valid for velocity dispersion dominated spherically symmetric systems but in practice, it is applied to estimate masses of a variety of dwarf galaxies (see discussions in \citealt{gonzalez-samaniego17} for details).

We have applied this equation to the line-of-sight velocity dispersion and circularized $r_{\rm eff}$ for our dwarf galaxies and compared them to the actual enclosed mass within the 3D half-light radii. \footnote{To get the 3D half-light radius we follow the observational approach and estimate it from the circularized, de-projected effective radius, $r_{\rm 1/2}=r_{\rm eff} \times \sqrt{b/a}\times 4/3$, where $r_{\rm eff}$ and the axis ratio, b/a, are calculated by GALFIT.} When used directly with $r_{\rm eff}$ from GALFIT, this approach tends to over-predict the mass within the 3D effective radius and it shows large variations between different sight-lines. We get a better agreement and no systematic offset when we use the same approach after the gas removal and subsequent relaxation (see the Appendix \ref{agasre}), which tends to make galaxies smoother. Furthermore, when we apply this equation to the actual 2D half-mass radii (instead of half-light) we recover the actual enclosed mass within the 3D half-mass radius to within 15\% when averaged over three orthogonal projections (consistent with the tests of this mass estimator on a lower mass FIRE-2 simulations by \citealt{gonzalez-samaniego17}). We therefore conclude that velocity dispersion and effective radius can indeed provide reasonable estimate of the enclosed mass but a larger number of measured systems are needed to reach a robust measure of the typical enclosed masses of UDGs.

\begin{figure}
 \includegraphics[scale=0.45]{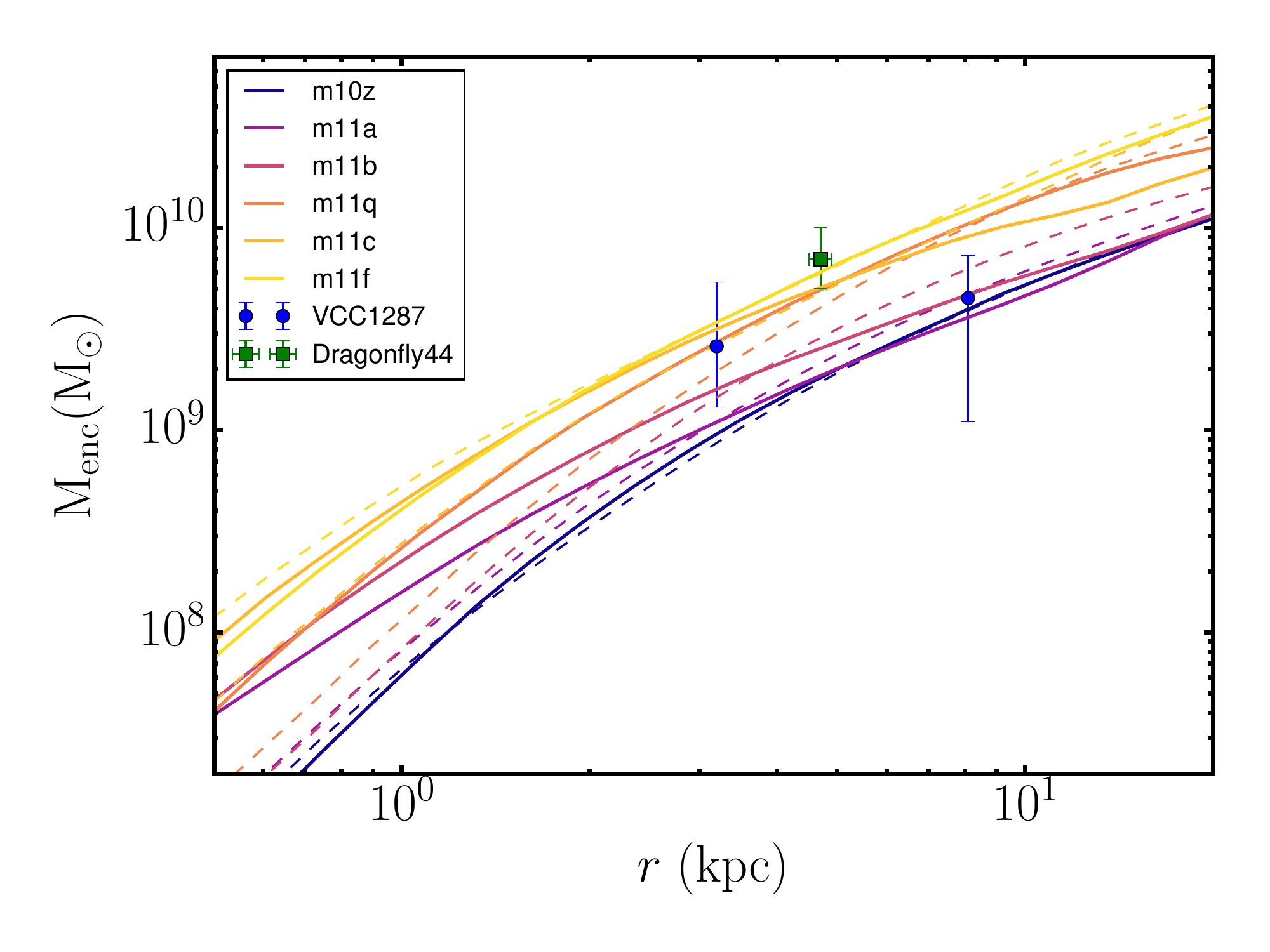}
\caption{Cumulative mass profiles of our simulated halos including stars and DM, compared to the observations of VCC 1287 ({\it blue circle}) \citep{Beas16a} and Dragonfly 44 ({\it green square}) \citep{vanD16}. Solid (dashed) lines show the mass profiles at the earliest (latest) quenching time when simulated galaxies match the UDG selection criteria and the range of absolute magnitudes from \citet{vanD15} (see Table \ref{UDGs}). While our dwarf galaxies/halos match the outer measurement of VCC 1287, our most massive galaxy/halo {\bf m11f} in the sample is a good match for Dragonfly 44.}
\label{Mencudg}
\end{figure}

In Figure \ref{Mencudg} we present the enclosed mass profiles of our simulated halos, including their stellar and DM components\footnote{While gas can be a non-negligible component in our galaxies, we leave it out of the enclosed mass calculation as we are comparing to red UDGs in clusters.}. Profiles are compared to the inferred values from the observations of \citet{Beas16a,vanD16}. The outer point from \cite{Beas16a}, already studied in Figure \ref{Menc}, is well matched by several of our lower mass dwarf galaxies suggesting that it is indeed a "quenched dwarf". 

The innermost point of VCC 1287 was calculated by the same authors with Equation \ref{Wolf}, with an additional assumption that GC velocity dispersion represents stellar velocity dispersion. This suggests a higher halo mass and it can be matched by four of our highest mass halos {\bf m11b}, {\bf m11q}, {\bf m11c} and {\bf m11f}. We note that using GC velocity dispersion instead of stellar velocity dispersion could be problematic because they only used 7 GCs, one of which is $r \gtrsim 3r_{\rm eff}$ away from the galactic center. However, even with this limitation our {\bf m11b} galaxy (with a $z=0$ halo mass of $4.4 \times 10^{10}\msun$) can match both mass measurements, suggesting that VCC 1287 formed in a dwarf mass halo. Future spectroscopic studies with long integration times are needed to constrain its actual stellar kinematics.

The mass measurement of Dragonfly 44 came directly from stellar kinematics and Equation \ref{Wolf}, and it is matched by the largest galaxy in our sample, {\bf m11f}. 
{\bf m11c}'s mass profile provides a marginal match, but only for the latest quenching time for which it is still identified as a UDG. At the quenching times when this galaxy satisfies the UDG criteria, its halo mass is only $\sim0.5-1\times 10^{11}\msun$, implying its total mass today $\sim 1\times10^{11}\msun$, assuming that its growth is insignificant after its infall. 


The number of GCs in Dragonfly 44 is $94^{+25}_{-20}$, much higher than inferred from its stellar mass and luminosity \citep{vanD16}.  Interestingly, according to relations in \citet{Harr13}, halos with $\sim94$ GCs have masses similar to that of {\bf m11f} at $z=0$.  If all of the GCs in this system formed at very early times and the galaxy's stellar mass growth was stopped at high redshifts, one should actually expect a very high number of GCs despite its low stellar and halo mass at quenching. This is because galaxies used for the \citet{Harr13} relation continue growing their stellar and halo masses to much later times, unlike quenched UDGs. This may also explain the finding of \cite{Amor16b} that the ratios between the number of GCs and stellar mass are much higher in some UDGs than the galaxies studied in \cite{Harr13}.

We have also checked the more massive galaxies presented in \cite{FIRE2}, {\bf m12z}, {\bf m12i} and {\bf m12c}, with $z=0$ halo masses $8.7\times 10^{11}$, $1.3\times 10^{12}$ and $1.4\times 10^{12}\msun$ respectively. They can only match the absolute magnitude and effective radius of Dragonfly 44 for $t_{\rm q}\lesssim 2\,{\rm Gyr}$, i.e. $z_{\rm q} \gtrsim 3.3$. At those times, their halo masses were around $10^{11}\msun$ and their inner masses matched Dragonfly 44. Hence, Dragonfly 44 can form in a halo similar to the MW's progenitor, but it has to be quenched very early. This means that at the time of quenching, Dragonfly 44 is not hosted by a highly "over-massive" DM halo. Its central halo mass, which appears unusually high for its estimated stellar mass, can be explained by its early formation time.

While forming a massive UDG requires very early quenching, we note that the Coma cluster was not fully formed so early in the structure formation process: the Coma cluster progenitor at $t=2$ Gyr is expected to have less than 5\% of its present-day halo mass \citep{Li07}. However, it is possible that the most massive UDGs ($M_\star\gtrsim 2\times 10^8\msun$) were quenched in group-mass progenitors that existed at high redshift. Indeed, semi-empirical constraints suggest that the majority of quiescent low-mass satellites in galaxy clusters today quenched in a group \citep{Wetz13}. The exact mechanism and the feasibility of such a scenario have to be explored with cosmological simulations that follow formations of galaxy groups or clusters.

Relatively early quenching of massive UDGs is also suggested by spectroscopic observations \citep{Maka15,Kado17,Gu17}. In particular, \citet{Gu17} inferred the stellar age and metallicity of Dragonfly 44 to be $8.9^{+4.3}_{-3.3}$ Gyr and  ${\rm [Fe/H]=-1.3^{+0.4}_{-0.4}}$, respectively. Our {\bf m11f} simulation has similar metallicity at $t_{\rm q}$ when it is a UDG but our age estimate is close to the upper range suggested by observations (see Table \ref{UDGs}). More massive simulated halos result in even older stellar ages for the allowed range of quenching times. However, we stress that uncertainties in age measurement are too large for an accurate determination of the quenching time of Dragonfly 44.

Our analysis therefore suggests that galaxies like Dragonfly 44 formed early ($z\sim 3$), in $\sim 10^{11} \msun$ halos and stopped growing after their infall into a cluster. These objects could continue accreting mass, forming stars and reaching much higher luminosities and halo masses by $z=0$ if they did not fall into a cluster. But due to the cluster influence, they should have stellar and halo masses significantly lower than those hosting ${\rm L}\star$ galaxies at $z=0$, since their formation and quenching likely took place in much lower mass halos. Overall, our simulations suggest that a majority of UDGs in clusters form in halos that span a relatively broad range of halo masses (${\rm a\; few} \times 10^{10-11} \msun$), with more massive UDGs forming in more massive halos. 


\begin{table*}
\centering
    \begin{tabular}{llllllllllllllll}
    \hline\hline
&&\multicolumn{2}{c}{\bf m10z}&\multicolumn{2}{c}{\bf m11a}&\multicolumn{2}{c}{\bf m11b}&\multicolumn{2}{c}{\bf m11q}&\multicolumn{2}{c}{\bf m11c}&\multicolumn{2}{c}{\bf m11f}\\
\hline\hline
\rule{0pt}{3ex} $\left \langle \sigma \right \rangle$  [km/s]&&       $21.4$& $\,_{20.1}^{22.6} $&      $21.6$& $\,_{19.5}^{24.4} $&      $23.8$& $\,_{23.1}^{24.1} $&      $31.2$& $\,_{27.0}^{36.6} $&      $29.5$& $\,_{27.2}^{33.4} $&      $47.7$& $\,_{39.0}^{54.2} $\\
    \hline
    \end{tabular}
    \caption{ Line-of-sight velocity dispersion $\left \langle \sigma \right \rangle(=\sqrt{\left<\sigma^2_{\rm los}\right>})$
    calculated at $t_{\rm q}$ when galaxies satisfy the UDG criteria and their g-band absolute magnitude is closest to $M_{\rm g}\sim -14.3$, the average absolute magnitude of UDGs in \citet{vanD15}. Velocity dispersions are measured in a cylindrical aperture with radius of 20 kpc, that includes vast majority of stars of each simulated UDG. The larger fonts show values averaged over three perpendicular directions, whereas the smaller fonts represent the maximum and minimum values of velocity dispersion as measured from three different perpendicular directions.
    For reference,  $\left \langle \sigma \right \rangle$  of VCC 1287 is $33^{+16}_{-10} {\rm km/s}$ \citep{Beas16a} and of Dragonfly 44 is $47^{+8}_{-6}  {\rm km/s}$ \citep{vanD16}.}
    \label{veldis}
\end{table*}

\subsection{Galaxy Expansion}
\label{galaxyexpansion}

One of the distinctive features of UDGs is their diffuseness. While we have already shown that stellar feedback leads to large effective radii and quenching of star formation can redden their colors, a large number of UDGs have been discovered in clusters, leading to a natural question: can satellite galaxies be further puffed-up with tidal heating and ram-pressure stripping? We explore the dynamical effects of gas removal in the Appendix \ref{agasre} and show that this mildly increases the size and reduces the surface brightness further.

Without invoking clusters, \citet{Yozi15}, \citet{Amor16} and \citet{Rong17} proposed that UDGs are diffuse because their progenitors have larger angular momenta compared to normal galaxies. In other words, they are the {\it high spin tail} of the galaxy population. Yet, high spin galaxies are also more likely to resemble disk-like \citep{Yozi15} rather than spheroidal structures as observed in UDGs\footnote{\cite{Yozi15} found their galaxies have high axis ratios but they only considered face-on images.}. We note (as discussed in \S~\ref{sec:simulations}) that most of our simulated UDGs have normal spin parameters and we find no clear differences between the UDGs forming in low and high spin halos \footnote{We note that unlike Dragonfly 44, two of our galaxies {\bf m11b} and {\bf m11f} do develop clear stellar disks at late times (without quenching). However, at those times they are not identified as red UDGs.}.
  
Observations also seem to contradict the "high spin tail" scenario. For example \cite{vanD16} showed that Dragonfly 44 is dispersion-dominated with no evidence of rotation and radial variations in the velocity dispersion. Similarly, \cite{Burk17} argued that from their axis ratios, UDGs are unlikely puffed-up disk galaxies, but are instead similar to dwarf spheroidal galaxies. 

Tidal stirring provides a possible pathway from gas rich dwarf irregulars with rotational support to gas free dwarf spheroidals through repeated tidal interactions with a massive host galaxy \citep{Maye01,Klim07,Klim09,Loka15}. But this mechanism may not explain the abundance of UDGs with high axis ratios observed in a variety of environments, from cluster, cluster outskirt, and group, e.g \cite{Mart16,vande16,vande17,Roma17a,Roma17b}. Furthermore such mechanism might not be needed to transform most of the dwarf irregulars as a large fraction of the isolated dwarfs are likely dispersion dominated \citep{Whee17}.


Stellar feedback can produce diffuse and spheroidal stellar distribution independent of host interactions. Through stellar migration and dynamical heating, feedback can decrease both the surface brightness and ellipticity of a dwarf galaxy while simultaneously increasing its effective radius \citep{Chan15,ElBa16}. One generic effect of stellar feedback in dwarfs with $M_{\rm h} \sim 10^{10-11}\msun$ is a cored stellar profile \citep{Read05,Stin13,ElBa16}, implying a flat central light profile and low Sersic index\footnote{The average Sersic indexes of our galaxies are $0.8\pm 0.4$.}. A flat surface brightness profile has been observed in one of the biggest UDGs in the Coma cluster \citep{vanD15b} and low Sersic indices ($\sim 0.8-0.9$) in UDGs were reported in various observations \citep[e.g.][]{Koda15,Roma17b}.  

If feedback is the major driver of their diffuseness, it is very natural to expect an abundant population of diffuse galaxies far from cluster centers and even in the field, which we will discuss in \S~\ref{implication}. \cite{Roma17a} and \cite{Mart16} found galaxies with large $r_{\rm eff}$ and low surface brightness even in under-dense regions, so cluster interactions are likely not an essential factor for their diffuseness, consistent with the scenario where feedback plays the dominant role in shaping the UDGs.

Furthermore, for feedback-driven radial migration, old stars experienced more feedback episodes than young stars, so older stars will migrate outside and young stars remain near the center. We should therefore expect mixed or even inverted age gradients in UDGs \citep{ElBa16}, i.e. stars far from the center might be older than the stars at the center, which could be observed in the future.

\subsection{Gas Removal and Quenching}
In order to quench dwarf galaxies, their gas supply needs to be truncated and their ISM gas also needs to be largely removed (or consumed) in order to stop their star formation. But the exact mechanism of gas removal from UDGs is still an open question. Tidal stripping is one possible mechanism, but it tends to remove weakly bound stars near the edges and reduce the sizes of the galaxies, making them more compact rather than more diffuse \citep[e.g.][]{Read06}. \cite{mowla17} did not find any signature of tidal stripping out to 4 $r_{\rm eff}$ in the Coma UDGs. However, \cite{Venh17} found elongated and distorted shapes of the largest UDGs ($r_{\rm eff}>3{\rm kpc}$) in the Fornax cluster, which may indicate the effect of tidal stripping, but the total contribution is unclear. Simulating the interaction between a cluster and a high spin dwarf galaxy, \cite{Yozi15} showed ram pressure stripping can efficiently remove the gas and quench the dwarf galaxy if it falls in at $z\sim 2$.

Our model does not specify the gas removal mechanism but assumes that feedback-expanded dwarfs have their gas reservoir removed along with the truncation of their gas supply (so called "strangulation") as they fall into clusters, enabling them to quench their star formation and turn into red UDGs. This can occur because hot cluster environment shuts down gas accretion in infalling satellite galaxies (e.g. \citealt{Kere05, keres09a, simha09, vandV17}) while their gas reservoir can be either removed by ram pressure of hot gas or by a feedback episode shortly after infall. Exact nature of gas removal and prevention of further gas accretion will be explored in future work. 

In both the major text and Appendix \ref{agasre}, we assume instant quenching, since we expect a short quenching time scale in the cluster environments. \citet{Yozi15} showed the cluster can quench the dwarf galaxy within 2 Gyr with ram pressure stripping. \citet{Wetz15} and \citet{Fill15} constrained the quenching times of similarly low-mass ($M_*\sim 10^8 \msun$) LG satellites to be $\lesssim 3$ Gyr. Quenching in more massive clusters should occur even faster and more efficiently.

\subsection{Implications for blue dwarf galaxies}
\label{implication}
Without accounting for quenching, none of our simulated galaxies end up as red UDGs at the present. However, even without quenching, at $z=0$ three of our simulated galaxies with $M_*\sim 10^8\msun$ ({\bf m10z, m11a, m11b}) have large effective radii, low surface brightnesses and $M_g > -16$ (as shown in Figure \ref{infallplot}), i.e. they satisfy most of the UDG criteria. These diffuse galaxies are much bluer than the red UDGs in \citet{vanD15} ($g-i<0.7$). This implies that there should be a significant population of blue UDG-like dwarfs in the field and at the cluster outskirts with $\mu(g,0)>23.5\;{\rm mag/arcsec^2}$ and $r_{\rm eff}>1$ kpc. According to Figure \ref{infallplot}, these blue UDGs have young stellar ages and are spheroidal (axis ratio $\sim 0.6$) and typically dispersion supported (as shown in \citealt{ElBa17b}).

Indeed, there are `blue' UDGs observed far from galaxy clusters or even in groups, e.g. \cite{Mart16}, \cite{Merr16}, \cite{Roma17b}, \cite{Truj17} and \cite{Shi17}. \cite{Pand17} studied one of the UDGs in the sample of \cite{Mart16} and found younger stellar populations ($\sim 3$ Gyr) and higher metallicity ($[{\rm Z/Z_\odot}]\sim -0.6$) than UDGs in cluster environments (e.g. \citealt{Gu17}).

A similar population of bluer UDGs was recently observed by \cite{Roma17a} outside of the over dense region of the galaxy cluster Abell 168. Compared to UDGs near cluster centers, UDGs in lower density regions have similar effective radii and surface brightnesses, but higher luminosities, bluer colors, and slightly higher stellar masses. Based on these properties, \cite{Roma17a} suggested that blue UDGs could be a low surface brightness extension of regular dwarf galaxies. These observations also showed that unlike other low redshift galaxies, the stellar mass distribution of UDGs peaks at $10^8\msun$, coincident with the mass range of the most efficient dynamical effect of stellar feedback (e.g. \citealt{Gove12,DiCi14,Chan15,ElBa16,Toll16}) and in a good agreement with our predictions for the properties of field UDGs. It is therefore clear that a large population of galaxies can remain diffuse in the field, owing to the effects of stellar feedback (also suggested in \citealt{DiCi17}).

\begin{figure}
 \includegraphics[scale=0.45]{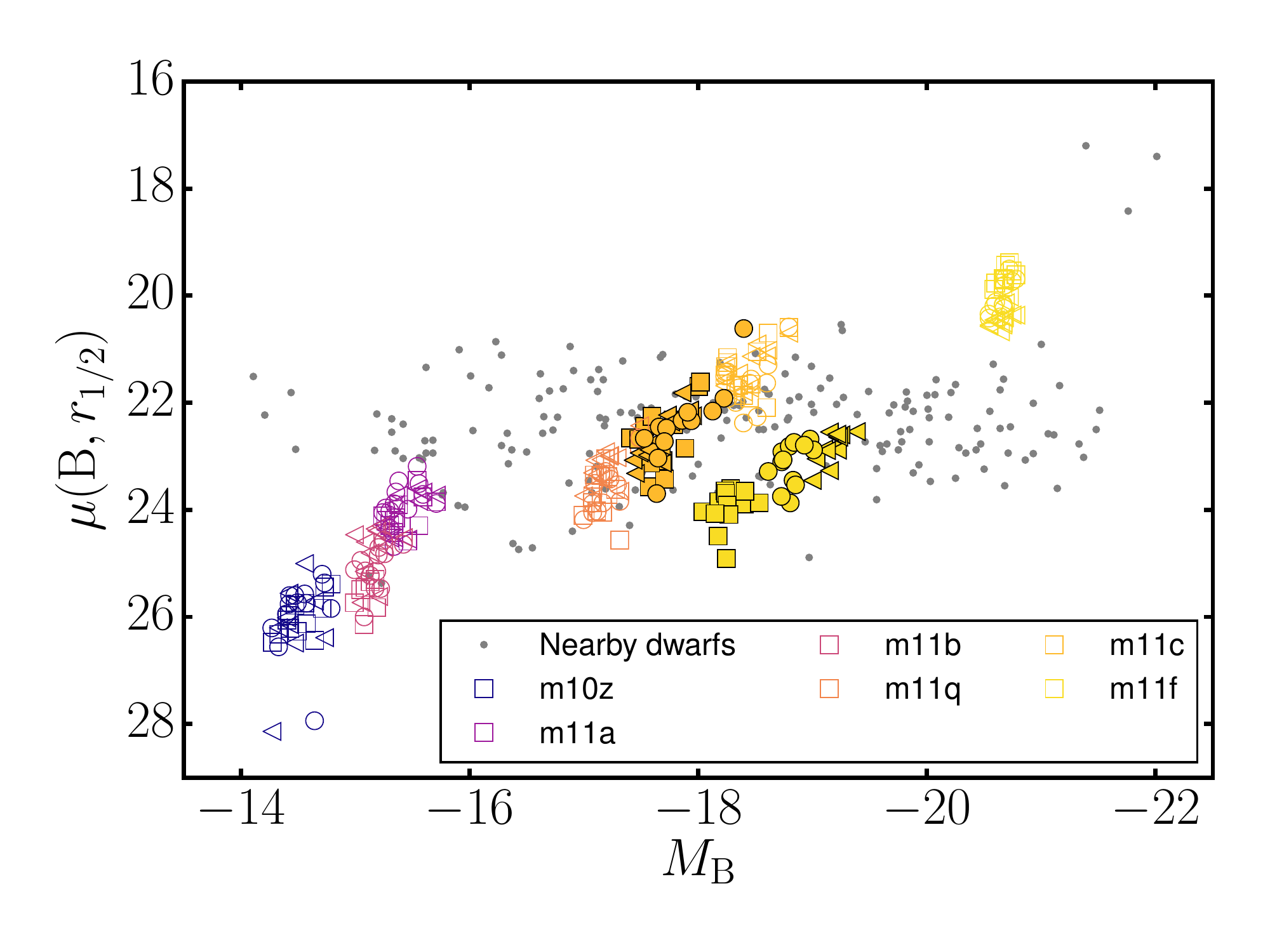}
\caption{B-band surface brightness, $\mu({\rm B}, r_{1/2})$, at a half light radius $r_{1/2}$ as a function of B-band absolute magnitude, $M_{\rm B}$, of our simulated galaxies for a number of simulation output times in the redshift range $z\sim 0-0.1$, compared to nearby galaxies from \citet{Jans00}. Empty symbols represent unattenuated values whereas solid symbols represent attenuated values. We show different lines of sight with different symbols, in the same manner as Figure \ref{surreff}. }
\label{bmag}
\end{figure}

Galaxies that, without quenching, reach much higher stellar masses by $z=0$ (e.g. {\bf m11q}, {\bf m11c} and {\bf m11f}) are too bright to be included in \cite{vanD15} sample. However one of them, {\bf m11q}, is still relatively diffuse at $z=0$ and could be an example of a more massive but more rare population of UDGs with $M_{\rm g}<-16$, $\mu(g,0)\sim 23-23.5$, $r_{\rm eff}\sim 1-5$ kpc and $g-i<0.8$. Indeed, several observations, e.g. \cite{Miho15,Roma17b}, found brighter examples of UDGs.

Finally, all of our galaxies are simulated as field dwarfs in cosmological simulations without the influence of a cluster, so at $z=0$ they should resemble field dwarf galaxies. \citet{ElBa16} showed that the effective radii of galaxies in FIRE simulations agree with those of the observed galaxies in NASA-Sloan Atlas \citep{Blan11}, resembling both the trend and scatter of the sample. In Figure \ref{bmag} we compare properties of our simulated galaxies with the sample of nearby field dwarfs from \cite{Jans00}. We consider our galaxies at $z\sim 0-0.1$ (to account for the star formation and feedback driven size variations) without passive aging, and, following the observations, measure their surface brightnesses $\mu({\rm B}, r_{1/2})$ at 2D half light radii $r_{1/2}$ in B-band.

Given that the observations are in B band, we calculate both the attenuated and unattenuated luminosities of star particles and estimate galactic luminosity, effective radii and surface brightness for both cases\footnote{We assume the gas to dust ratio from \citet{Bouc85} scaled by metallicity, the SMC-like dust extinction curves from \citet{Pei92} and use the method from \cite{Hopk05} to calculate the dust attenuation of stellar light.} and show results in Figure \ref{bmag}. For low-mass dwarfs, attenuated and unattenuated values are almost the same so we only show attenuated values for our higher mass dwarfs, $M_* \sim 10^{9}\,\msun$, where differences are significant. The figure shows that our galaxies provide a reasonable match to the observed nearby field dwarfs, although simulated $M_*\sim 10^8\msun$ dwarfs tend to be lower surface brightness than the dimmest dwarfs in the observed sample. We note that potential complex selection effects in the observed sample are not taken into account. \cite{Jans00} noted that the relative completeness at a given luminosity of their sample, especially at low surface brightness, is not well characterized, leaving out a potentially large population of low-surface brightness dwarfs such as the ones in our simulations. Much lower surface brightness galaxies indeed exist in the under-dense environment near clusters \citep[e.g.][]{Roma17a}.

\begin{figure}
 \includegraphics[scale=0.45]{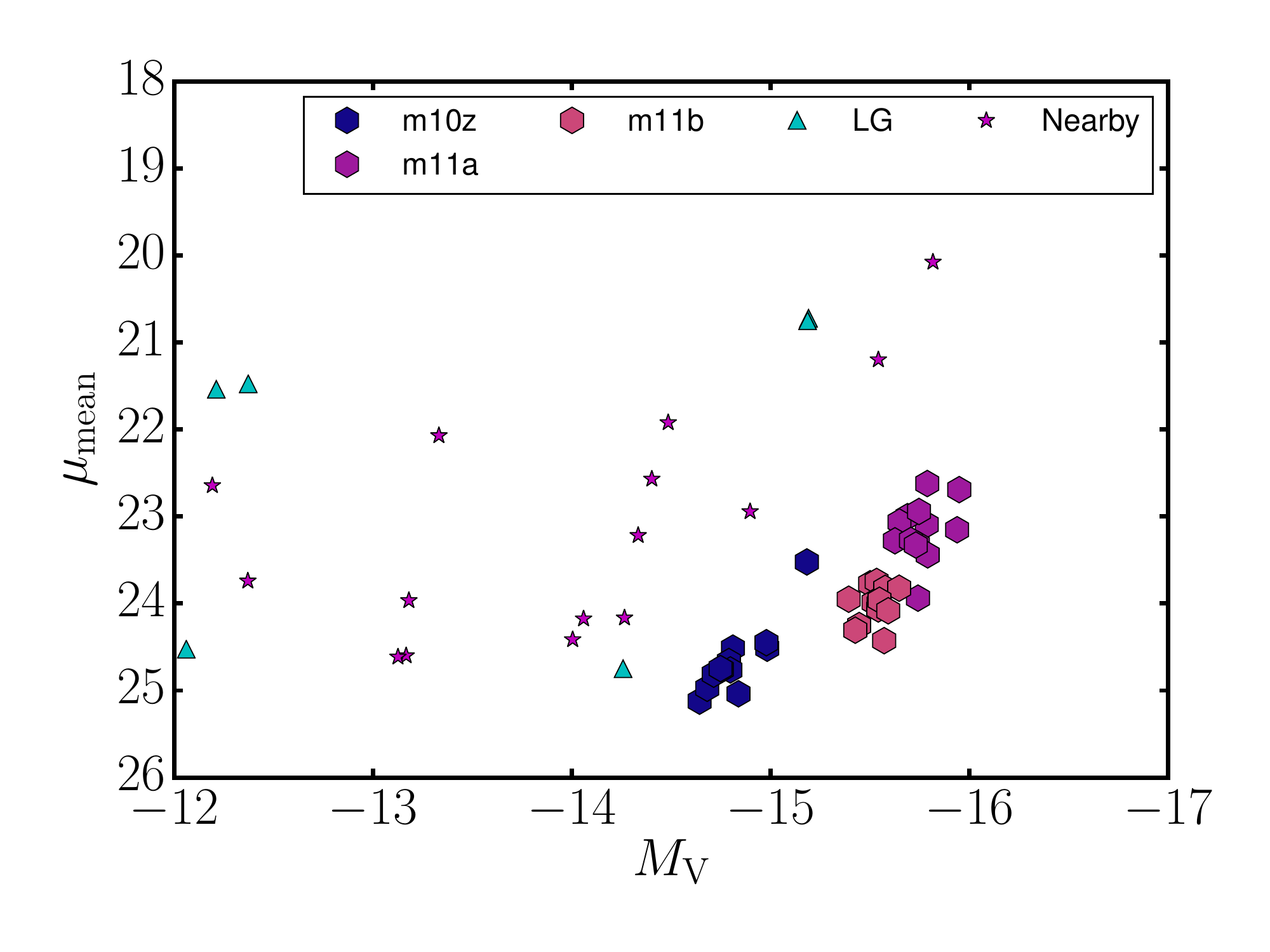}
\caption{Mean V-band surface brightness within the effective radius, $\mu_{\rm mean}$, viewed along x direction as a function of V-band absolute magnitude $M_{\rm V}$  for three of our simulated galaxies at $z\sim 0-0.1$, compared to the LG and nearby galaxies from \citet{McCo12}. The nearby galaxies are defined as isolated galaxies that do not belong to any major galaxy grouping but are still within 3 Mpc of the Sun.}
\label{vmag}
\end{figure}

In Figure \ref{vmag}, we compare simulated dwarf galaxies with observed dwarfs in the Local Group and nearby regions \citep{McCo12}, who use different photometric bands and probe galaxies to lower magnitudes than the \citet{Jans00} sample. Following \citet{McCo12}, we measure the mean surface brightness within the circular isophote defined by the half light radius. We do not consider passive aging and attenuation, and show the results at $z=0-0.1$ to account for the occasional bursts of star formation. The figure shows three of our simulated galaxies whose absolute magnitude overlaps with \citet{McCo12} sample. Our galaxies resemble the trend of the higher mass end of the nearby dwarfs, although simulated galaxies have somewhat lower surface brightnesses.  Field dwarfs are slightly higher surface brightness (i.e. more compact) than UDGs of the same absolute magnitude. At V-band magnitude $M_{\rm V} \sim -14.5$ to -15.5, most of the field dwarfs are diffuse with $r_{\rm eff}\gtrsim 1\,{\rm kpc}$  and have V-band effective surface brightnesses $<\mu_{\rm V}>_e \sim 21-23.5\,{\rm mag/arcsec^2}$, while our simulated dwarfs have $<\mu_{\rm V}>_e \sim 22.5-25 \,{\rm mag/arcsec^2}$.

While our current sample is too small for a detailed statistical comparison, it does not produce any high-surface brightness analogues in the relevant mass range $M_*\sim 10^7-10^9 \msun$, which could indicate excessive expansion by feedback.
We note, however, that the observational sample is limited by the small survey volume and multiple examples of isolated galaxies with much lower surface-brightness (typically by 2-3 ${\rm mag/arcsec^2}$) than the \citet{McCo12} sample exist \citep[e.g.][]{Dalc97,Roma17b,Bell17} outside of LG.  In the nearby universe, deep optical follow-ups of HI detected objects (already tested in e.g. \citealt{tollerud15}) or hostless transients due to novae or SNe \citep{Conr15}, or a large CCD survey \citep[e.g.][]{Dalc97} could be used to uncover an even larger number of blue UDGs. 

Our results suggest that UDG-type surface brightness is the dominant outcome of galaxy formation in low-mass halos that host galaxies with stellar masses $\lesssim 10^8\msun$ ($M_{\rm h}\lesssim 5\times10^{10}\msun$), with a caveat that our simulated sample currently contains only a small number of galaxies. While our lower mass examples appear to have lower surface brightness than local observed dwarfs, we note that the full population of low redshift field galaxies has not yet been properly characterized at very low surface brightness, and that typical observed samples are biased toward high surface brightness. For example, \citet{Huan12} found that about half of the HI selected dwarf galaxies in their survey, which have typically low surface brightness, do not have a counterpart in the SDSS spectroscopic survey, suggesting that current surveys miss a significant fraction of such objects. More careful analysis of these sources revealed that a large fraction of these galaxies have properties similar to blue UDGs \citep{Leis17}. Our simulated galaxies that remain diffuse at $z=0$ with $\mu(g,0) > 23.5 \,{\rm mag/arsec^2}$, ({\bf m10z}, {\bf m11a} and {\bf m11b}), are all gas rich with corresponding gas fractions, $f_{\rm gas}=m_{\rm HI}/(m_{\rm HI}+m_*)$, of 0.57, 0.55 and 0.9, respectively and could represent such HI-rich UDGs.

\cite{DiCi17} proposed a formation scenario for field UDGs very similar to ours: feedback-driven gas outflows affect stellar profiles in the central regions in the same way as DM core formation, and produce diffuse low surface brightness dwarf galaxies. They considered isolated galaxies in cosmological simulations in the Numerical Investigation of a Hundred Astrophysical Objects (NIHAO) project \citep{Wang15}, showing that their dwarf galaxies with $M_{\rm h}\sim 10^{10-11}\msun$ at $z=0$ can match the surface brightness of observed UDGs, similar to our finding for the field UDG population, despite differences in stellar feedback models and hydrodynamical methods\footnote{While \citet{DiCi17} only studied blue field UDGs with simulated isolated or central galaxies fully evolved to $z =0$, we also account for the effects of quenching, which enables us to more directly address the formation mechanism of red UDGs, commonly observed in clusters.}.

Given the burstiness of star formation and resulting outflows in FIRE simulations of dwarf galaxies \citep{Mura15}, our simulations make specific predictions for  blue UDGs formed by stellar feedback. They should have: (a) a range of sizes at fixed stellar mass depending on where they are in their burst cycles, (b) mixed or even inverted age and metallicity gradients \citep{ElBa16}, and (c) sizes and velocity dispersions that correlate with their recent star formation history \citep{ElBa17a}.

\section{Conclusions}
\label{conclusions}
We study the origin of UDGs using FIRE-2 cosmological simulations of field dwarf galaxies with halo masses $M_{\rm h}(z=0)\sim 10^{10-11}\msun$. Our earlier work with the FIRE simulations \citep{ElBa16,Chan15} showed that in this halo mass range, stellar feedback and associated changes in the gravitational potential are the most effective in dispersing both DM and stellar populations in the inner halo. In addition, newly-formed stars can inherit the velocity of the star forming gas cloud pushed by an outflow episode and further expand the stellar population. Here we show that these mechanisms lead to a diffuse quasi-spherical stellar distribution with surface brightness and overall properties comparable to observed UDGs.

We then assume that star formation and growth of progenitors of UDGs stop (i.e. galaxies "quench") during infall into a cluster of galaxies as a combination of tidal and gas stripping processes prevents fresh gas supply and removes the existing gas. To mimic this quenching, we artificially stop star formations of UDG progenitors at a cosmic time $t_{\rm q}$ and passively evolve their stars to $z\sim0$ according to a stellar population synthesis model (FSPS; \citealt{Conr09}). Finally we generate synthetic images and use GALFIT \citep{Peng02} to estimate their central surface brightness and effective radii. Our main findings are summarized below:

\textbullet All of our simulated galaxies with $M_*\sim 10^7- 10^8\,\msun$ are diffuse and spatially extended ($\mu(g) \gtrsim 23.5\, {\rm mag/arcsec^2}$ and $r_{\rm eff} \gtrsim 1.25\,{\rm kpc}$). These galaxies are typically hosted in halos with $M_{\rm h}\sim 3\times10^{10}-1\times10^{11}\,\msun$ at their quenching times.

\textbullet  The dynamical effects of stellar feedback produce UDGs even without taking into account cluster influence. Gas removal can help to further expand the galaxies, as shown in Appendix \ref{agasre} but this effect is likely of secondary importance.

\textbullet DM halos of our simulated galaxies have a typical distribution of spin parameters, suggesting that formation of UDGs does not require high spin halos.

\textbullet Red UDGs require quenching of star formation. Our simulations indicate that typical UDGs are dwarf galaxies quenched over a wide range of times ($t_{\rm q}>3\;{\rm Gyr}$). Simulated analogs of observed red UDGs that form in halos $M_{\rm h}\sim 2\times10^{10}-1\times10^{11}\msun$ halos can be quenched over a broad time interval $t_{\rm q} \sim 5-11\;{\rm Gyr}$, i.e. redshift range $z\sim 0.3-2$.

\textbullet
The most massive red UDGs ($M_\star\gtrsim 2\times 10^8\msun$) in our simulations require earliest quenching. Our higher mass halos can host red UDGs if their star formation and growth is quenched at very early times, $t_{\rm q}\sim 3-5\;{\rm Gyr}$, i.e. at $z_{\rm q}\sim 2-3$. At the time of quenching, the host halo mass of our most massive simulated UDG, similar to e.g. Dragonfly 44, is around $10^{11}\msun$. 

\textbullet Colors of red UDGs are approximately independent of quenching time as galaxies quenched later (i.e. with a younger stellar population) typically have higher metallicity. This prevents an accurate estimation of quenching time from $g-i$ color. 

\textbullet  Galaxies with  $M_*\lesssim 10^8\msun$ remain diffuse even at $z=0$ but have relatively blue colors. We predict that diffuse galaxies with bluer colors ($g-i<0.8$) are prevalent in the field. Our galaxies at $z=0$ match the magnitude-surface brightness relations of some samples of nearby galaxies, but have lower surface brightness than the LG sample from \cite{McCo12}. While our sample is small and statistics are limited, this raises an interesting prospect that there is a large number of undiscovered low surface brightness galaxies in the Universe.

\textbullet Given the UDG formation process in our simulations, the size and velocity dispersion of `blue' UDGs at a fixed mass should correlate with their recent star formation history.

\textbullet Our galaxies are not hosted in over-massive halos ($M_*/M_{\rm h}\lesssim 10^{-4}$) at quenching, but instead have a stellar-to-halo mass ratios similar to observed dwarfs, $M_*/M_{\rm h}\sim 10^{-3}$. 

\textbullet  The enclosed masses of our simulated galaxies can match the measured masses of observed UDGs in clusters, if we assume that the growth of their central enclosed mass stopped when they were quenched. Even if such halos had evolved in isolation outside of clusters, we showed that in most cases their enclosed DM mass on these scales would remain unchanged, owing to halo growth largely by ``pseudoevolution''. Owing to a broad range of formation times of these objects, inferred halo masses at z=0 can therefore lead to misleading conclusions about their host halo masses. This is especially important for the most massive UDGs whose dense central regions formed at very early times.

\section{ACKNOWLEDGEMETS}
We thank Aaron Romanowsky, Arianna Di Cintio, Timothy Carleton, and Viraj Pandya for helpful discussions. TKC was supported by NSF grant AST-1412153. DK was supported by NSF grants AST-1412153 and AST-1715101 and the Cottrell Scholar Award from the Research Corporation for Science Advancement. AW was supported by a Caltech-Carnegie Fellowship, in part through the Moore Center for Theoretical Cosmology and Physics at Caltech, and  by NASA through grants HST-GO-14734 and HST-AR-15057 from STScI. Support for PFH was provided  by  an  Alfred  P.  Sloan  Research  Fellowship, NASA  ATP  Grant  NNX14AH35G,  and  NSF  Collaborative  Research Grant \#1411920 and CAREER grant \#1455342. CAFG was supported by NSF through grants AST-1412836, AST-1517491, AST-1715216, and CAREER award AST-1652522, and by NASA through grant NNX15AB22G. KE was supported by a Berkeley graduate fellowship, a Hellman award for graduate study, and an NSF graduate research fellowship. Support for SGK was provided by NASA through Einstein Postdoctoral Fellowship grant number PF5-160136 awarded by the Chandra X-ray Center, which is operated by the Smithsonian Astrophysical Observatory for NASA under contract NAS8-03060. MBK acknowledges support from NSF grant AST-1517226 and from NASA grants NNX17AG29G and HST-AR-13888, HST-AR-13896, and HST-AR-14282 from the Space Telescope Science Institute, which is operated by AURA, Inc., under NASA contract NAS5-26555. The simulation presented here used computational resources granted by the Extreme Science and Engineering Discovery Environment (XSEDE), which is supported by National Science Foundation grant no. OCI-1053575, specifically allocation TG-AST120025. 

\bibliographystyle{mn2e}
\bibliography{mn-jour,mybib}

\appendix
\section{GALFIT with sky background}
\label{appendix:background}
\begin{figure}
 \includegraphics[scale=0.45]{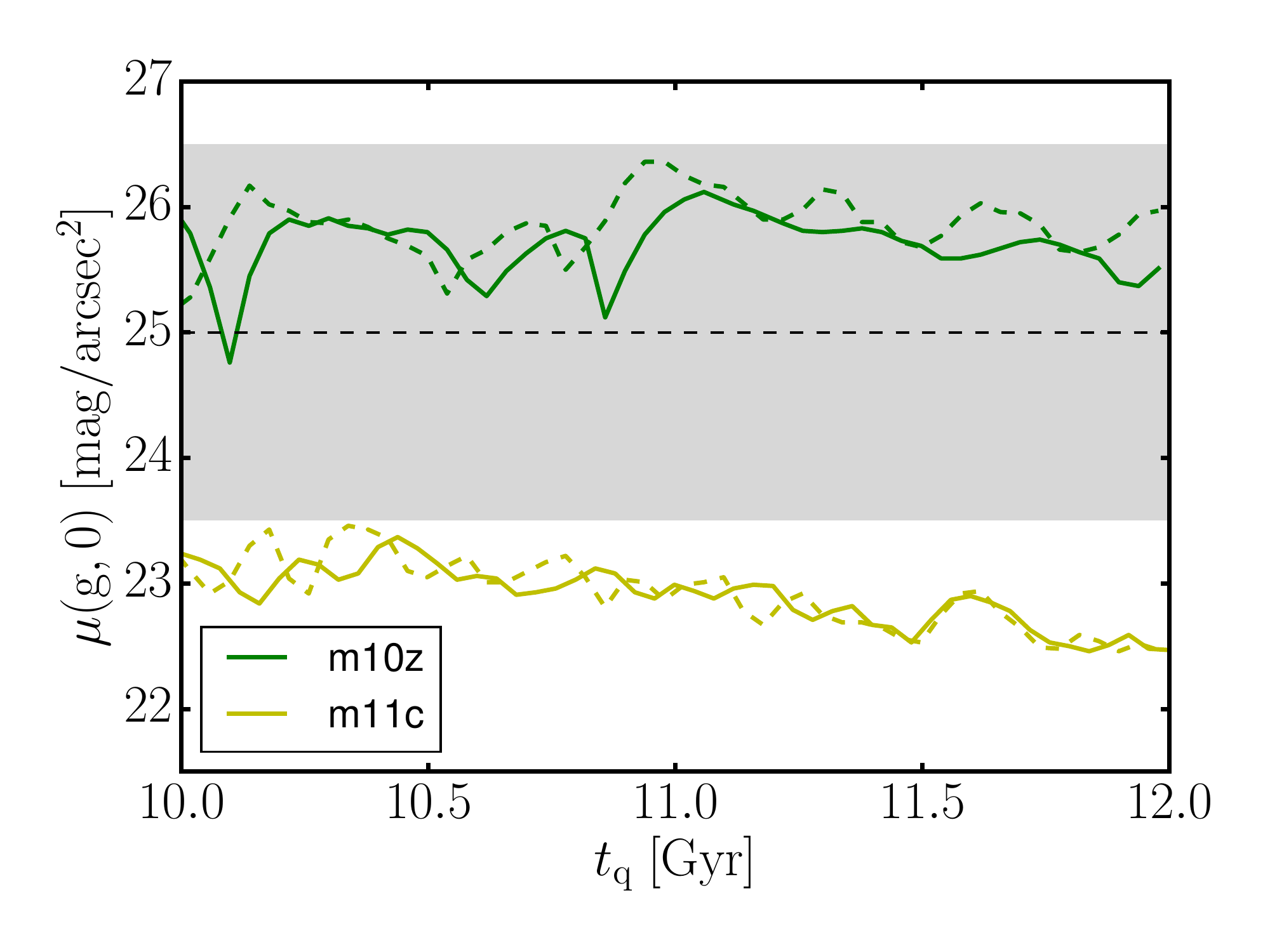}
  \includegraphics[scale=0.45]{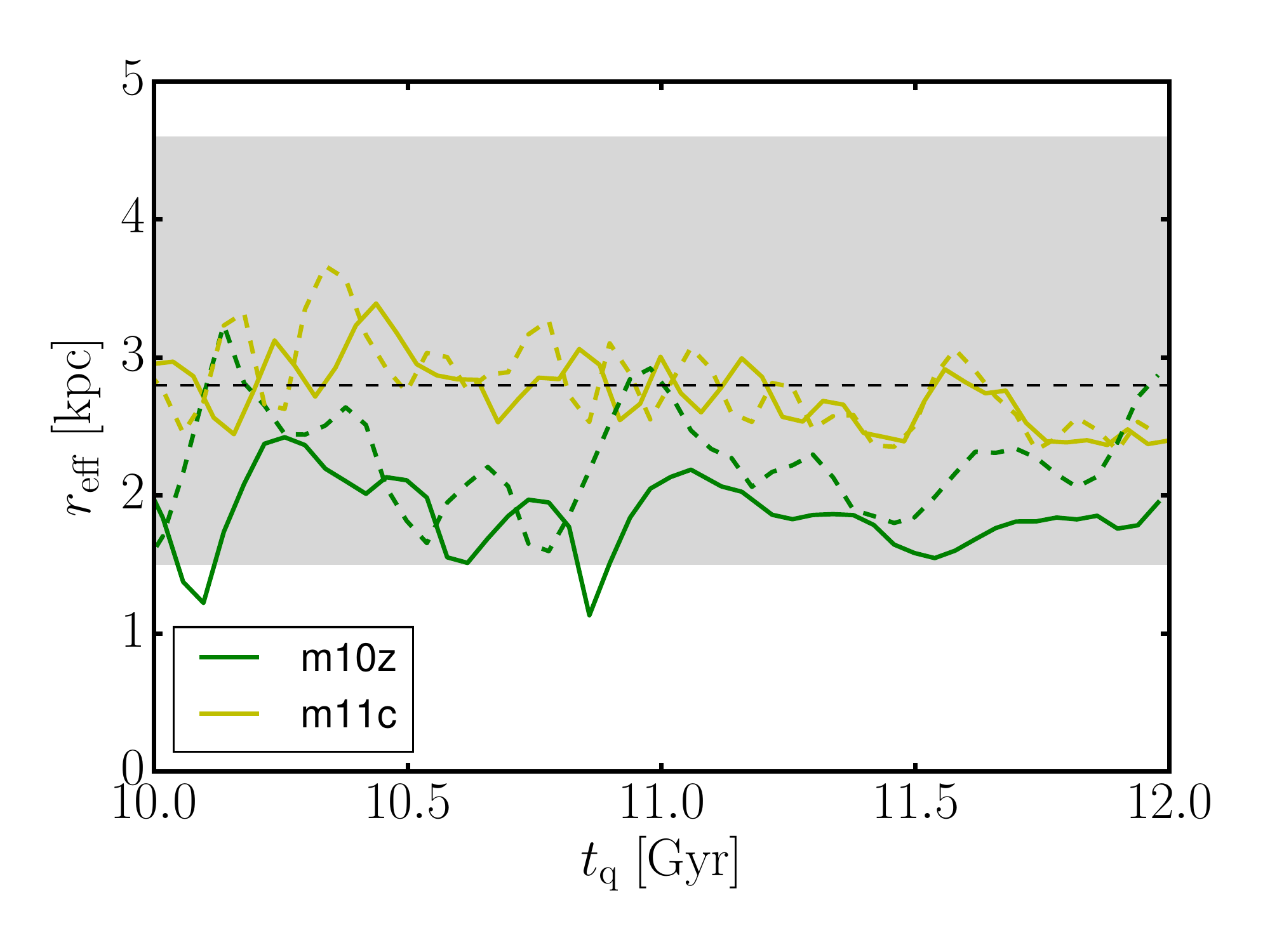}
\caption{Time evolution of central g-band surface brightnesses $\mu(g, 0)$ and effective radii of two of our simulated galaxies with a sky background (dashed) and without (solid) in their {\it g}-band images. The sky background is a random noise with the averaged surface brightness around 26 ${\rm mag/arcsec^2}$. $\mu(g,0)$ is obtained with GALFIT. Sky background and galaxy are fitted simultaneously. Stars are passively aged to $z=0$ and dust attenuation is not considered. Including a typical sky background has only a small effect on the estimated values.}
\label{musky}
\end{figure}

Realistic galaxy image contains sky background in addition to galactic light, so a sky subtraction is required. A proper subtraction is not trivial for galaxies with surface brightnesses comparable to the background, e.g. UDGs. To gauge the potential impact of sky background on the estimated properties of UDGs, we generate simulated galaxy images along with stochastic sky backgrounds whose average surface brightness is $\sim 26\,{\rm mag/arcsec^2}$. Then we estimate their central surface brightness and effective radius with a two-component fit, assuming Sersic profiles for galaxies and tilted flat planes for sky backgrounds. Figure \ref{musky} shows that the differences in central surface brightness and effective radius with and without sky backgrounds are small, even when the sky is, on average, brighter than the galaxy, illustrating the robustness of the fitting. The fitted $r_{\rm eff}$ differs from Figure \ref{infallplot} because here we use {\it g}-band instead of {\it g} + {\it i} images. 

\section{Dynamical effect of Gas Removal}
\label{agasre}

Most of the observed red UDGs are detected in galaxy clusters and were probably quenched through interactions with the host cluster. They do not show signatures of tidal interaction \citep{mowla17} but it is possible that their gas was removed by ram pressure stripping. To simulate the effect of gas removal on a dwarf galaxy, we test a simple toy model introduced in \citet{ElBa17a}: all gas particles instantaneously receive 1000 ${\rm km/s}$ velocity boosts at a given infall/stripping time. Then we evolve the galaxy to $z=0$ (by continuing the run in a fully cosmological environment) and estimate its properties with GALFIT as described in the main text.

Since the gas velocity after the kick is much higher than the escape velocity, all of the gas is quickly removed and the galaxy is quenched after around 100Myr. \citet{ElBa16} tested this method and concluded that the effect is almost identical to instantaneously taking out all of the gas particles. Fast moving particles also affect the surrounding gas so the galaxy can never accrete new gas and gets quenched. While in our default approach we only passively quench star formation and do not allow star particles to move after quenching, here they can freely move and adapt to a new and shallower gravitational potential. 

\begin{figure}
 \includegraphics[scale=0.41]{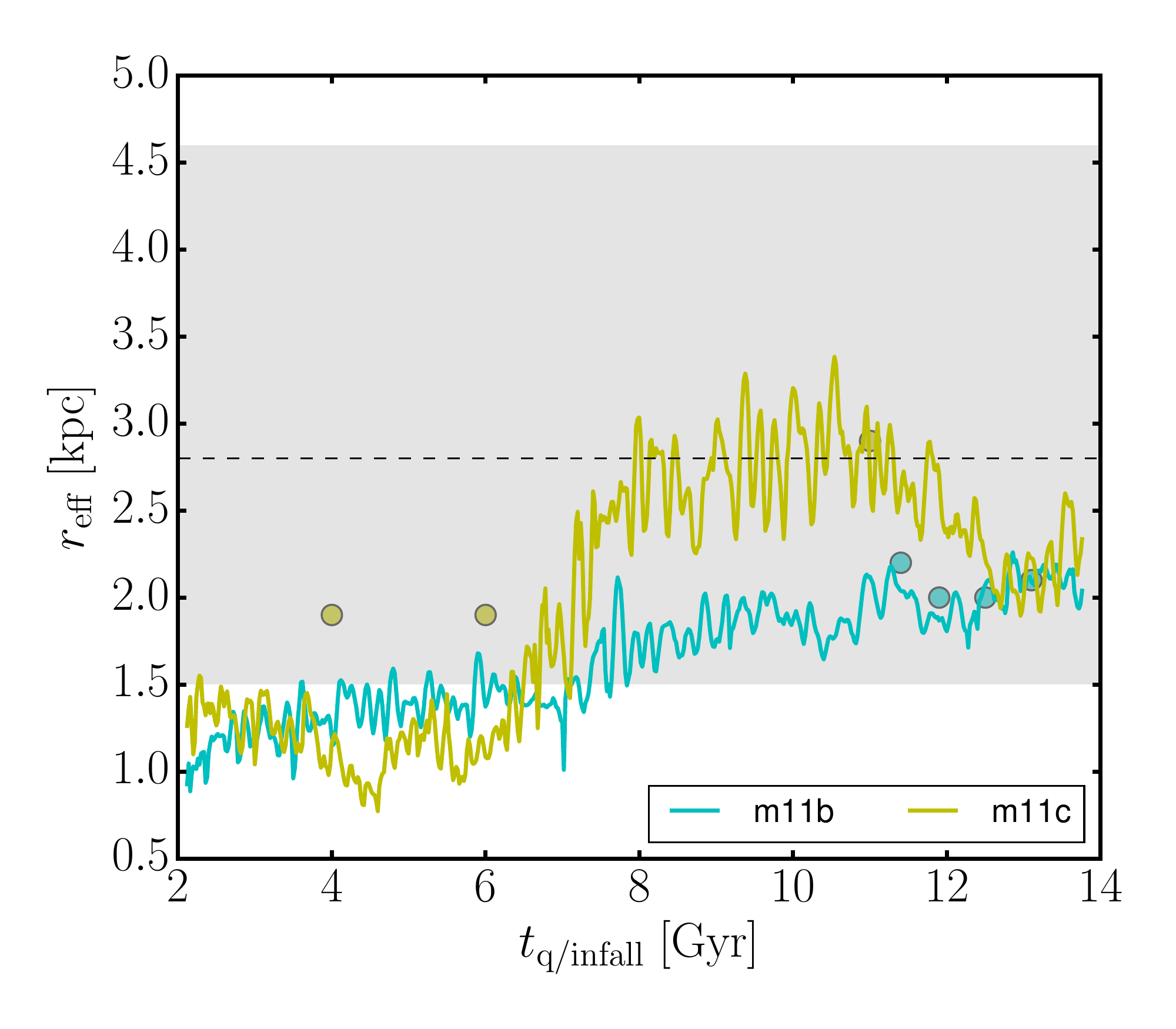}
 \vskip -12pt
 \hskip 4pt
 \includegraphics[scale=0.4]{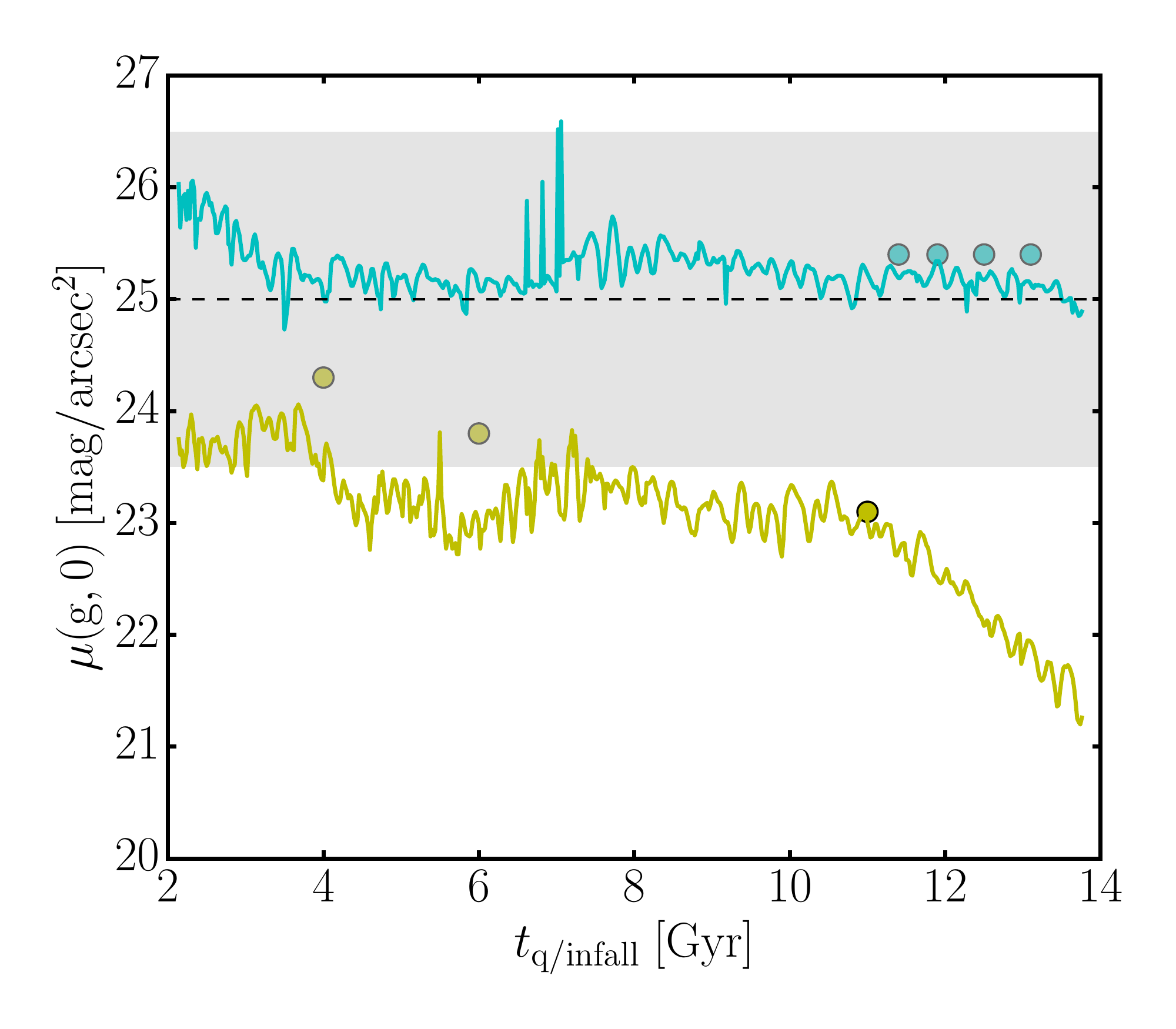}
  \vskip -12pt
  \hskip 7pt
 \includegraphics[scale=0.41]{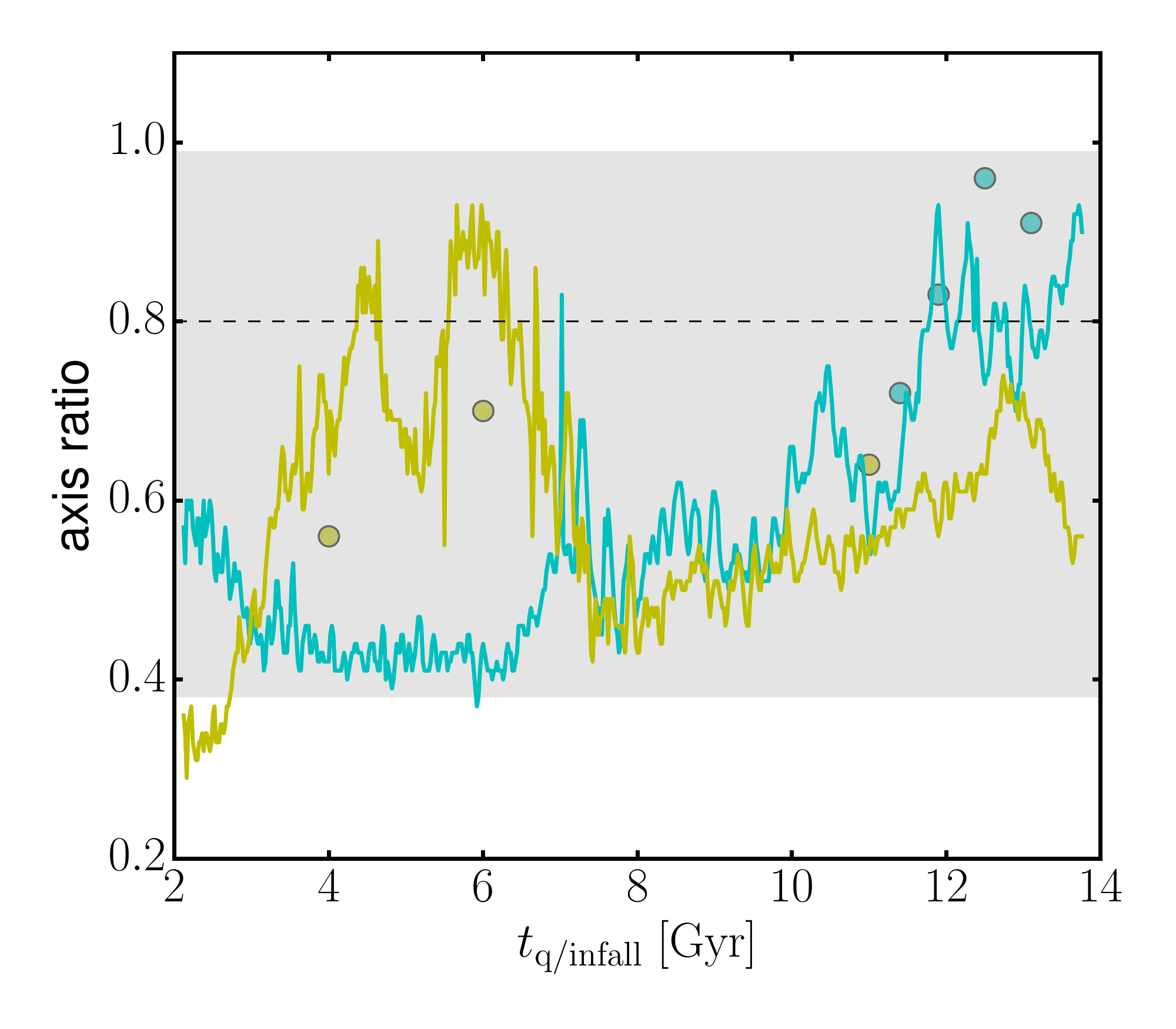}
  \vskip -12pt
\caption{Evolution of the effective radius, central surface brightness and axis ratio for two of our simulated galaxies. Lines show values from passive evolution scenario described in the main text, whereas points show values from gas-stripped {\bf m11b} and {\bf m11c} dynamically evolved to $z=0$. Lines are functions of quenching time $t_{\rm q}$, while individual points indicate several different times for which we apply our ram pressure approximation, mimicking the effect of hot cluster gas during the infall. Galaxies are typically quenched shortly ($\sim 100{\rm Myr}$) after gas stripping.}
\label{gasstripfig}
\end{figure}

Figure \ref{gasstripfig} shows that the effect of gas removal is small, since dynamical relaxation after gas stripping induces only a slight increase in size and a slight drop in central surface brightness, while there is no clear systematic effect on the axis ratio. The additional dynamical effects of gas stripping (compared to the fiducial model) therefore only help making our simulated galaxies slightly more diffuse in the relevant mass range. These results also show that galaxy sizes do not evolve much after ram pressure stripping and remain largely "frozen" in time motivating our approach of passively evolving galaxies after the quenching time. Overall both galaxies stay within or outside of the observational range of UDGs even when gas stripping is applied.

\begin{figure}
 \includegraphics[scale=0.45]{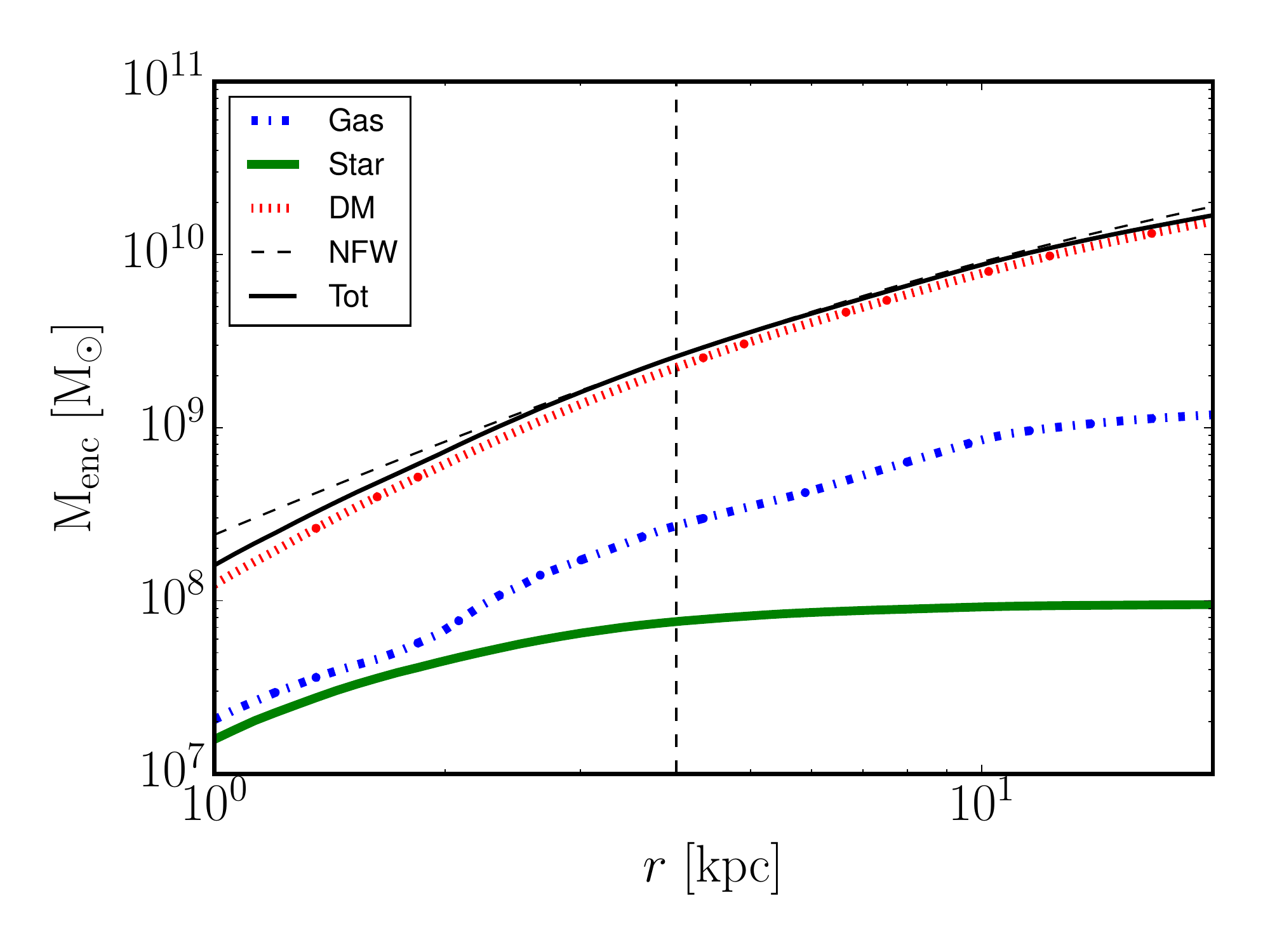}
 \includegraphics[scale=0.45]{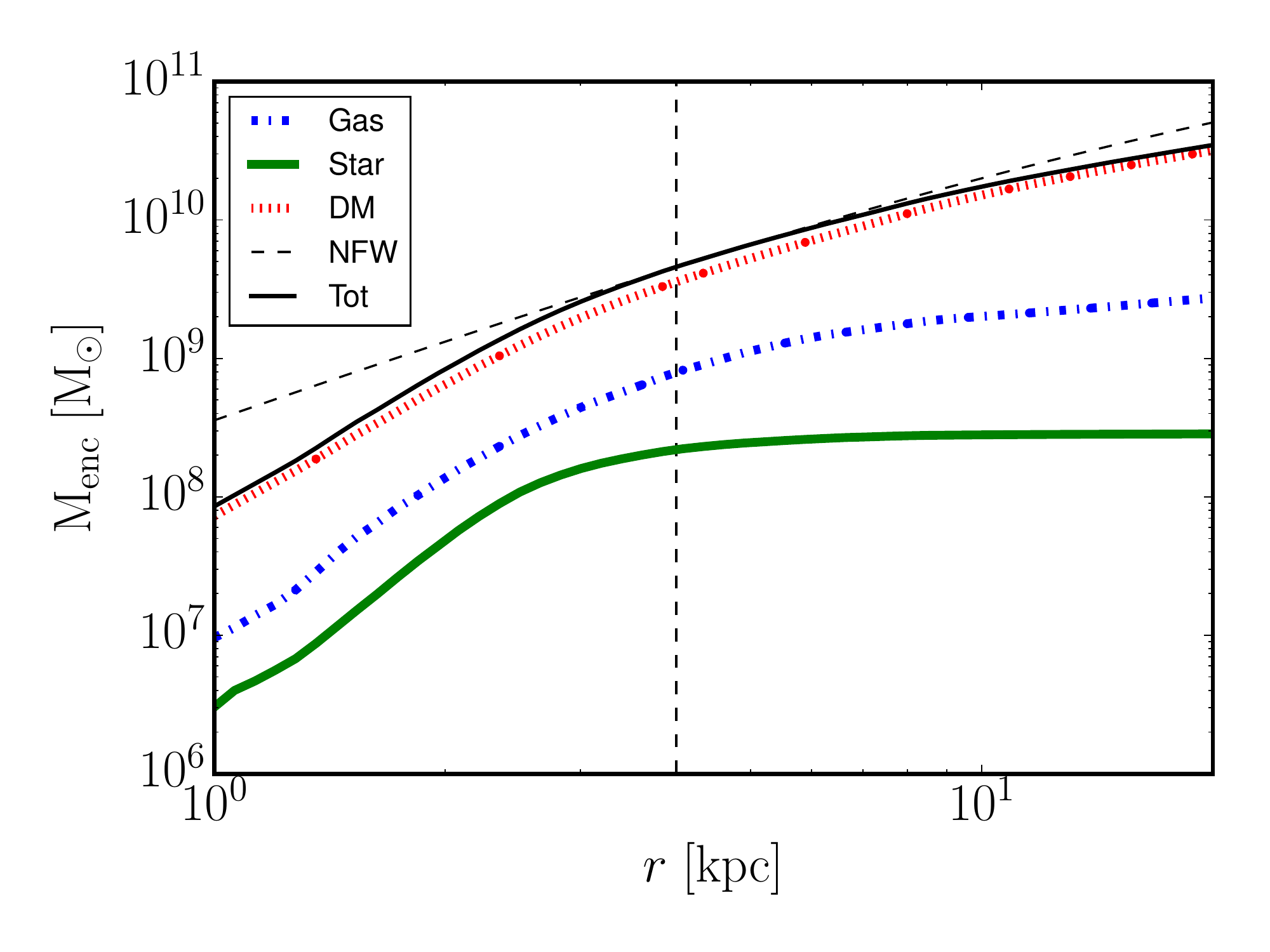}
\caption{Cumulative gas (blue;dash-dotted), star (green;solid), DM (red;dotted) and total (black;solid) mass profiles of {\bf m11b} at t=11 Gyr (Upper) and {\bf m11c} at t=5 Gyr. Dashed vertical lines show the effective radii and black dashed curved lines show the NFW profiles whose enclosed masses within the effective radii match our halos. DM dominates enclosed mass throughout the halo for both galaxies.}
\label{Mencgsdm}
\end{figure}

Relatively weak effect of gas removal is not entirely surprising given that galaxies are already largely spheroidal and dark matter dominated. While the gas dominates baryonic component, its gravitational influence is much weaker than that from the dark matter. This is illustrated in Figure \ref{Mencgsdm} where we show the enclosed mass profiles for stars, gas and dark matter in the inner halo for the two galaxies for which we apply quenching. Profiles are shown at characteristic times for which passively evolved counterparts correspond to the observed red UDGs. Dark matter dominates at all radii.

\section{Effect of resolution}
\label{relimit}
To evaluate the effect of resolution, we compare two of our galaxies, {\bf m10z} and {\bf m11c}, with their higher resolution versions, presented in \citet{FIRE2} (these have particle masses $m_{gas}=260\msun$ and $m_{gas}=2.1\times 10^3\msun$, respectively). The high resolution versions have particle masses eight times smaller and softening lengths twice shorter than those shown in Table \ref{SIC}. While their halo masses do not change with resolution, their stellar masses at $z=0$ drop by $40\%$ and $60\%$ in {\bf m10z} and {\bf m11c} respectively.

\begin{figure}
  \includegraphics[scale=0.45]{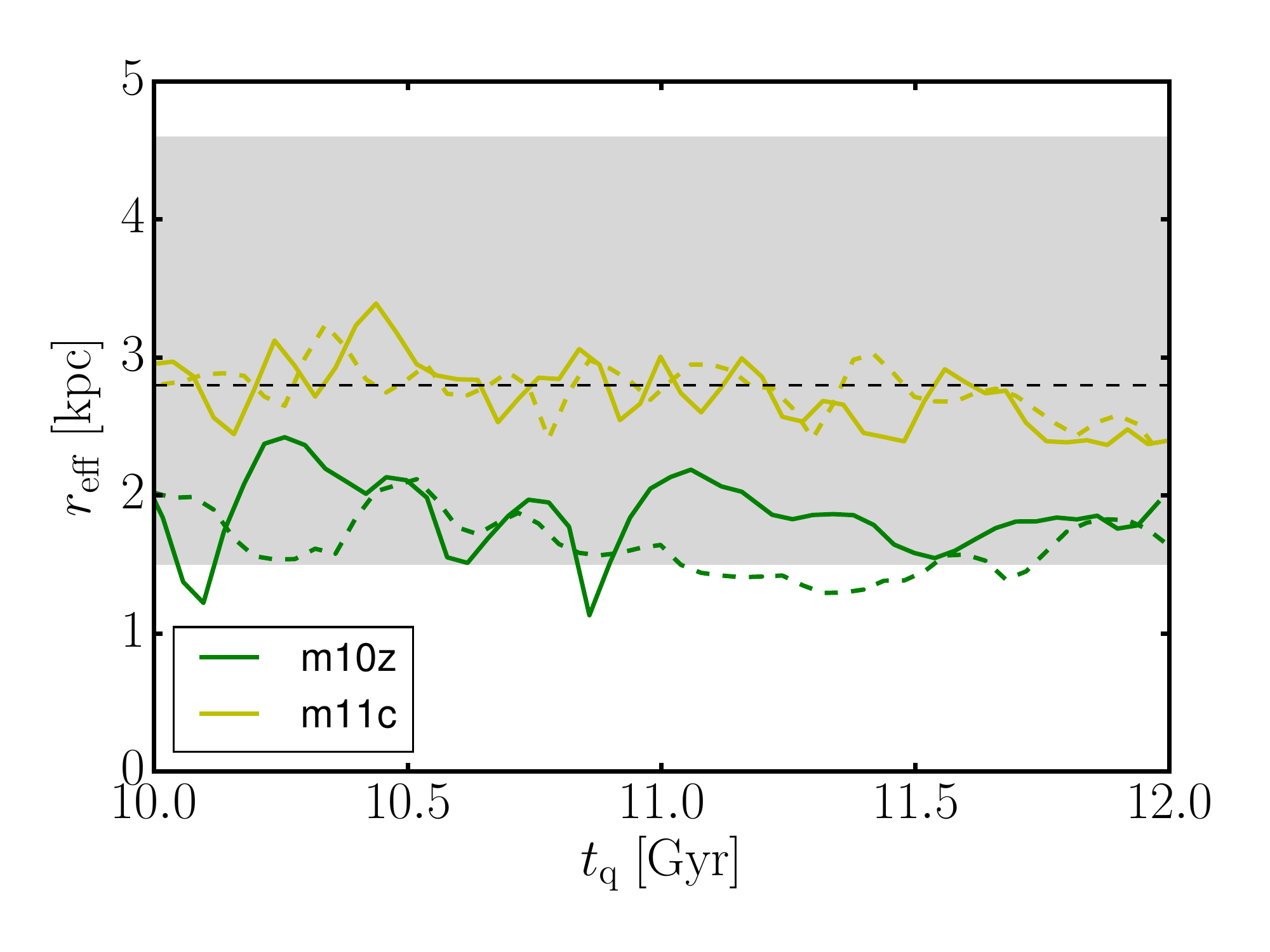}
   \includegraphics[scale=0.45]{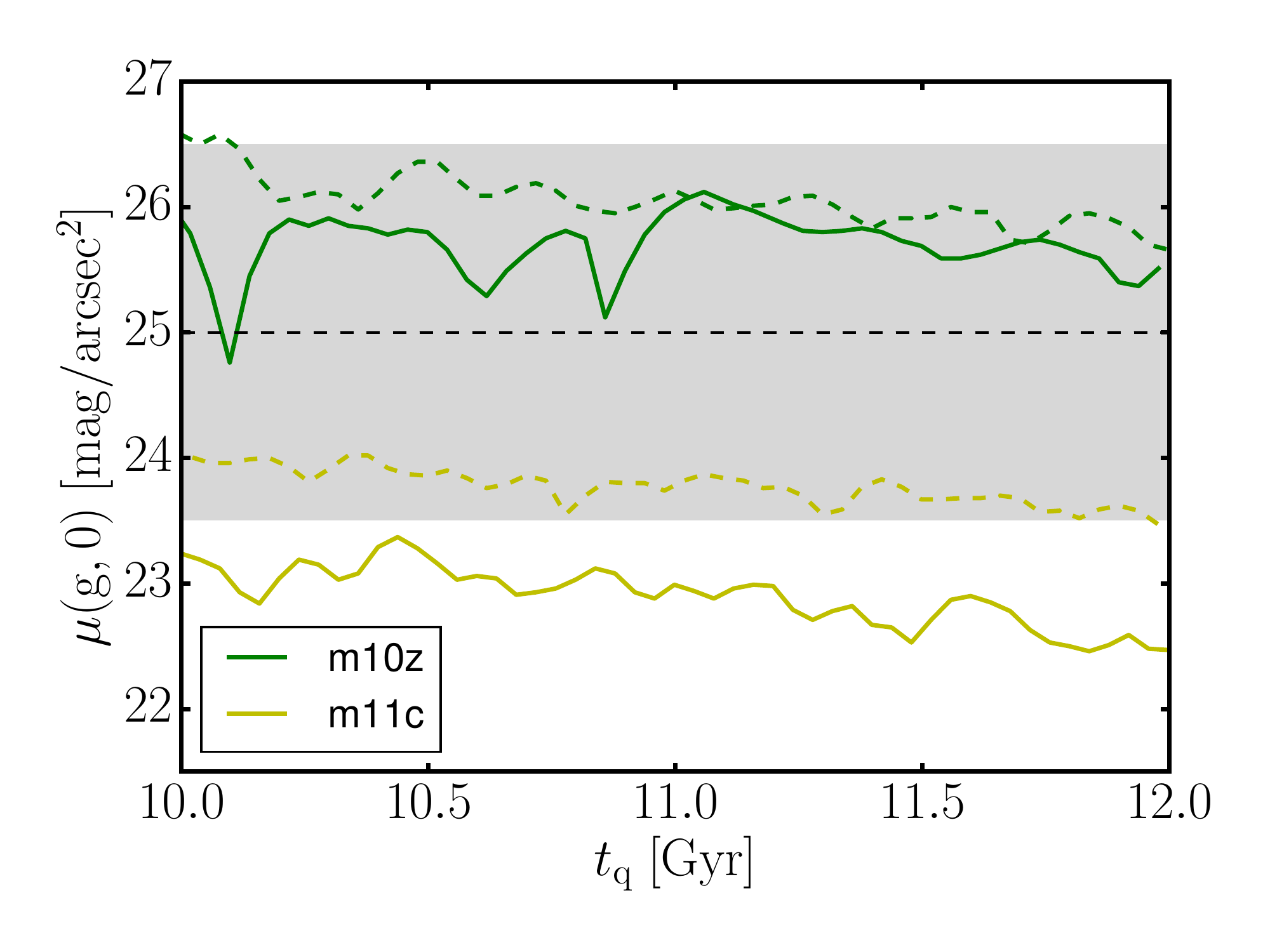}
  \includegraphics[scale=0.45]{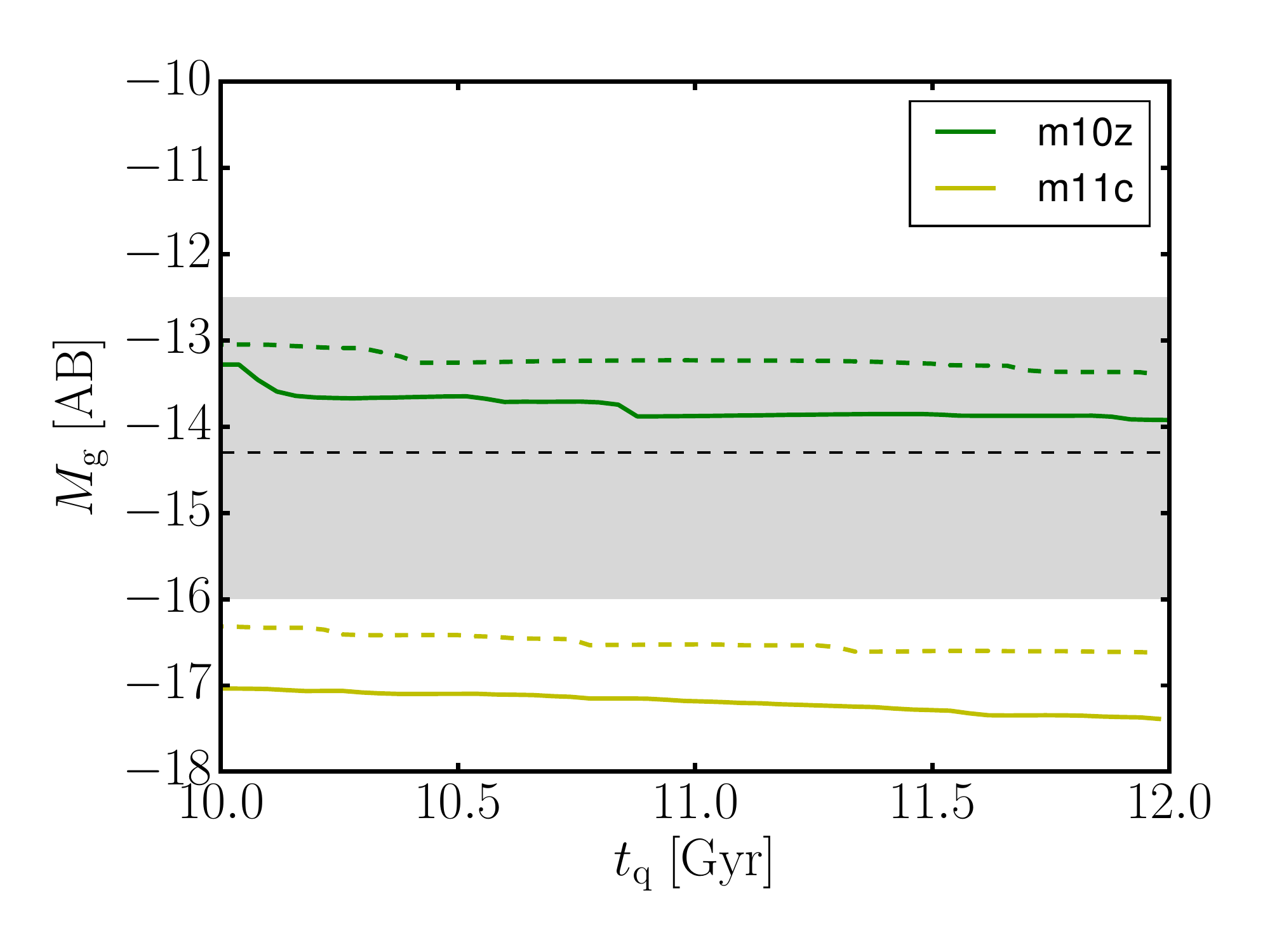}
\caption{Time evolution of central g-band surface brightness $\mu(g, 0)$ and effective radius for the two simulated galaxies with a higher resolution (dashed) and with the fiducial resolution (solid). Values are obtained from {\it g}-band images using GALFIT.}
\label{muhv}
\end{figure}

Figure \ref{muhv} shows the difference in central surface brightness, effective radius and {\it g}-band magnitude between two resolutions. The sizes of the galaxies are not sensitive to resolution, but their {\it g}-band magnitudes decrease by $\sim 1$, as expected from their smaller stellar masses. Their surface brightness also drops accordingly. These changes are consistent with what we expect from stellar mass difference between lower and higher resolution galaxies. We therefore conclude that while resolution can affect the stellar masses of our simulated galaxies, the surface brightness and size at a given stellar mass are largely not affected by resolution.

\section{GALFIT modeling with different resolution images}
\label{appendgalfit}

\begin{figure}
 \includegraphics[scale=0.45]{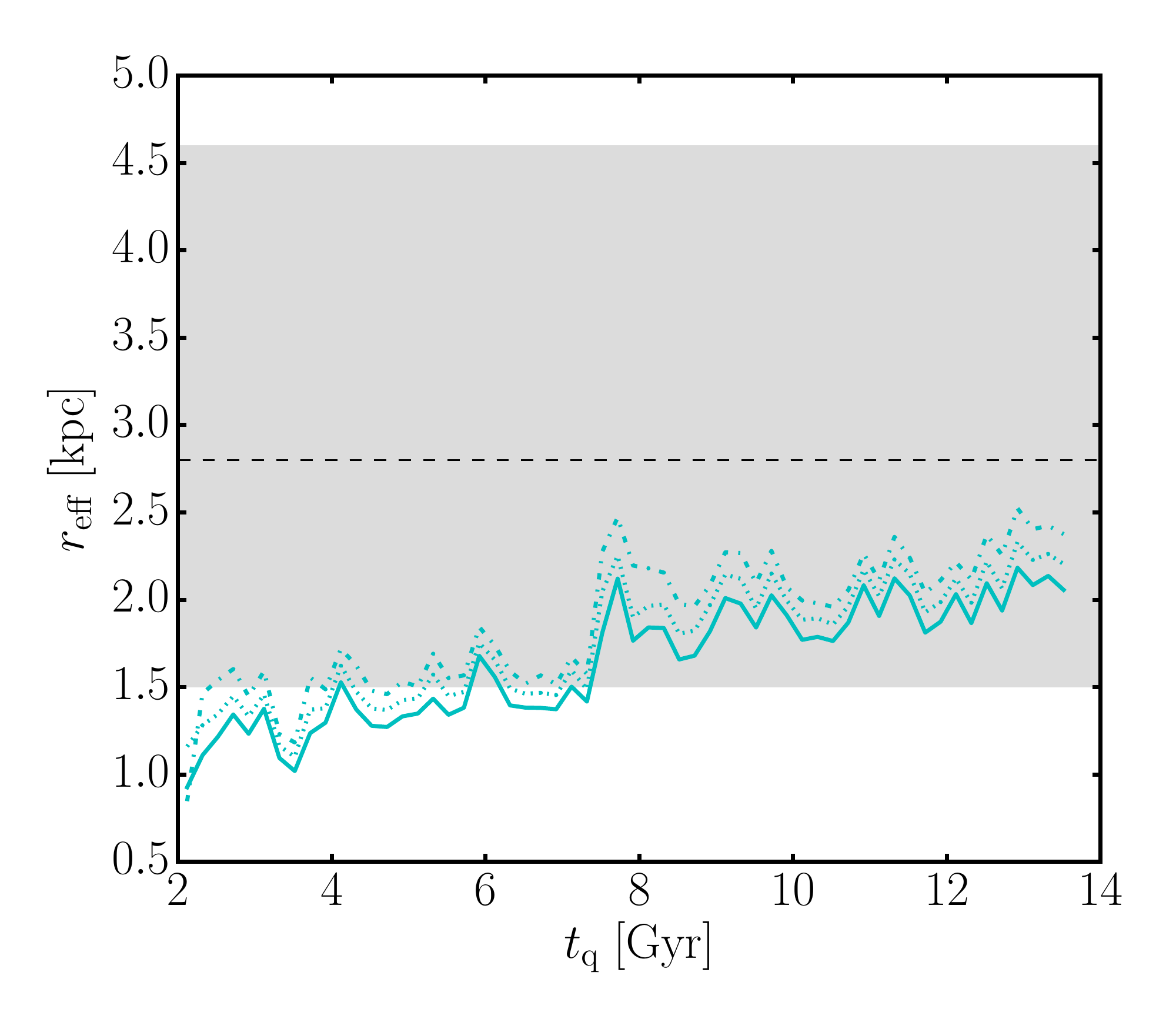}
 \vskip -5pt
 \includegraphics[scale=0.45]{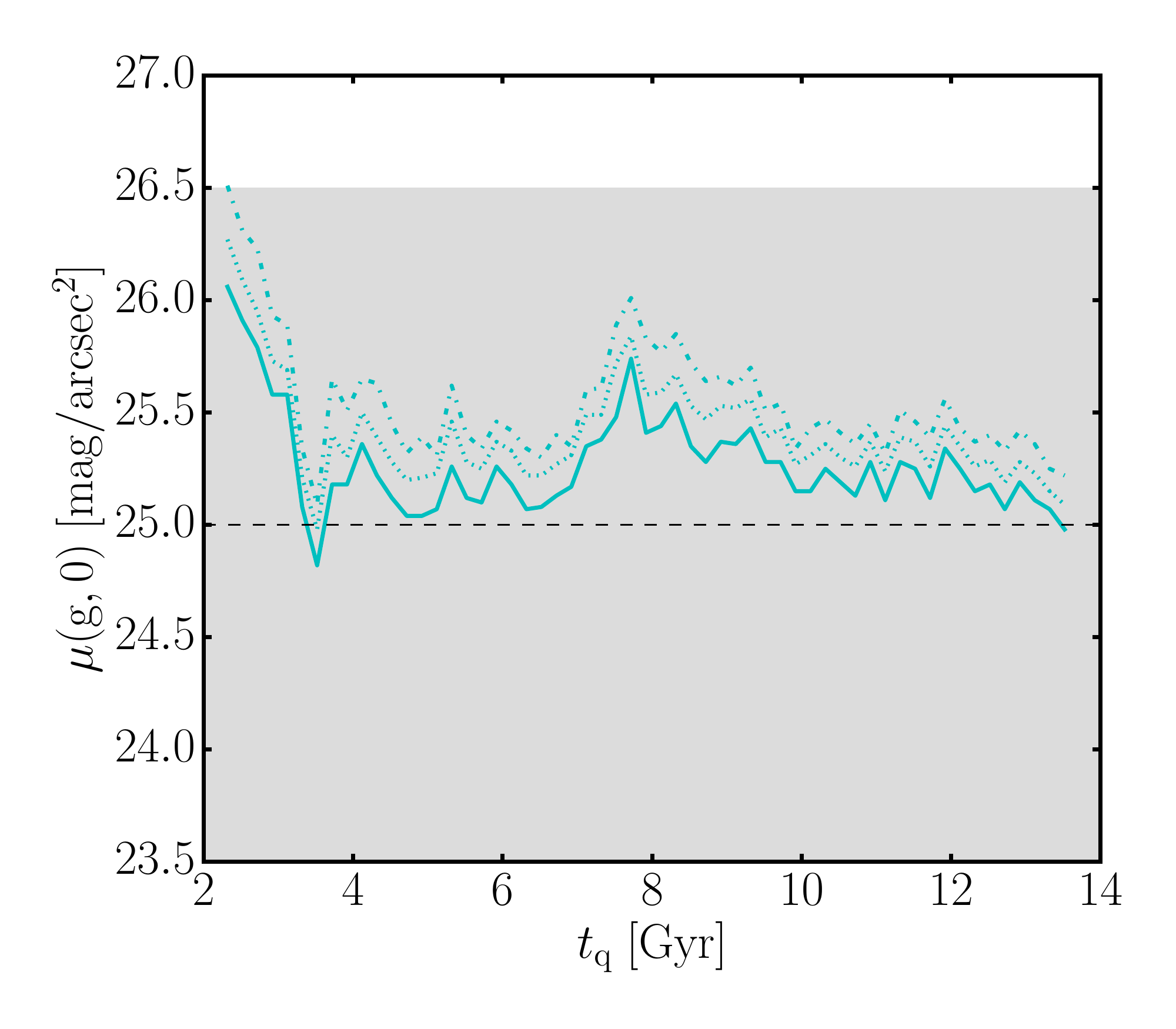}
  \vskip -5pt
\caption{Time evolutions of the effective radius and surface brightness of {\bf m11b} as calculated by GALFIT from 40 kpc x 40 kpc g band images with uniform-size pixels each with 400 pc (dashdot), 200 pc (dot) and 40 pc (solid) on a side (i.e. 100x100, 500x500 and 1000x1000 pixels).}
\label{resogalfit}
\end{figure}

Stellar particles in simulated galaxies have well determined positions and represent a population of stars with "radii" (i.e. gravitational softening) that is typically much finer than typical resolution from observations. To mimic a range of point spread functions (PSF) of different telescopes and a range of distances at which one observes UDGs, we vary the pixel size of our GALFIT images (i.e. the resolution of the 2-D projection of stellar properties) of our {\bf m11b} galaxy from 40 pc to 400 pc, and show results in Figure \ref{resogalfit}. This approximately spans the range of PSF between the Hubble Space Telescope and the Canada-France-Hawaii Telescope (CFHT) at the distance of the Coma cluster (i.e. 0.08 to 0.8 arcsec for the assumed distance of 100 Mpc).  For the tested range, there is a small systematic increase in effective radii and a slight decrease in surface brightness in low resolution images, but the change is smaller than the short term time variations of these properties.

\bsp

\label{lastpage}

\end{document}